\documentstyle[fleqn,psfig]{l-aa}

\newcommand{\ms}{M$_{\odot}$ }

\newcommand{\mss}{M$_{\odot}$}
\newenvironment{eqnum}[1]{\begin{center} \vspace{-1.2cm} \hfill (#1) 
\vspace{0.1cm}}{\end{center}}

\newcommand{\eg}{\rm e.g.\ }
\newcommand{\etal}{\rm et al. }
\newcommand{\ie}{\rm i.e.\/ }

\newcommand{\cf}{cf.\/ }

\renewcommand{\baselinestretch}{1.}

\begin{document}

\thesaurus{20( 08.01.1; 08.05.3; 08.16.4;  09.01.1; 11.01.1)}

\title{New theoretical yields of intermediate mass stars}

\author{L.B. ~van den Hoek\inst{1} and M.A.T. ~Groenewegen\inst{2} }

\institute{Astronomical Institute ``Anton Pannekoek'',
   Kruislaan 403, NL 1098 SJ Amsterdam, 
   The Netherlands
\and
   Max-Planck-Institut f\"{u}r Astrophysik, Karl-Schwarzschild-Stra{\ss}e 1, 
   D-85748 Garching, Germany}    

\date{Received date; accepted date}

\offprints{L.B. van den Hoek (bobby@astro.uva.nl)}

\maketitle
\markboth{New theoretical yields of intermediate mass stars}{}









\begin{abstract}

We present theoretical yields of H, $^{4}$He, $^{12}$C, $^{13}$C,
$^{14}$N, and $^{16}$O for stars with initial masses between 0.8 and 8
M$_{\odot}$ and initial metallicities Z = 0.001, 0.004, 0.008, 0.02, and
0.04.  
We use the evolutionary tracks of the Geneva group up to the early 
asymptotic giant branch (AGB) in combination with a synthetic thermal-pulsing 
AGB evolution model to follow in detail the chemical evolution and mass loss 
up to the end of the AGB including the first, second, and
third dredge-up phases. Most of the relations used are metallicity
dependent to make a realistic comparison with stars of different
initial abundances. The effect of Hot Bottom Burning (HBB) is included
in an approximate way.

The free parameters in our calculations are the mass loss scaling
parameter $\eta_{\rm AGB}$ for stars on the AGB (using a Reimers law),
the minimum core mass for dredge-up M$_{\rm c}^{\rm min}$, and the
third dredge-up efficiency $\lambda$. As derived from previous extensive
modeling, $\eta_{\rm AGB}$ = 4, M$_{\rm c}^{\rm min}$ = 0.58 \mss, and
$\lambda = 0.75$ including HBB are in best agreement with observations of 
AGB stars both in the Galactic disk and Magellanic Clouds.

The influence of specific model assumptions and adopted parameter
values on the resulting AGB yields is examined and compared with
earlier theoretical work.  We compare the abundances predicted during
the final stages of the AGB with those observed in planetary nebulae
in the Galactic disk and show that the model with the
aforementioned parameters is in good agreement with the observations.
The metallicity dependent yields of intermediate mass stars presented in
this paper are well suited for use within galactic chemical evolution
models.

\end{abstract}


\section{Introduction}

Presumably all main sequence stars with initial masses between $\sim
0.9$ and $\sim$8 \ms pass through a double-shell burning phase at the
end of their lifetime, also referred to as the asymptotic giant branch
(AGB) phase.  During this phase, intermediate mass stars lose most of
their envelope mass while they contribute substantially to the
interstellar abundances of He, C, N, and s-process elements (\eg
Renzini \& Voli 1981; Iben \& Renzini 1983; Dopita \& Meatheringham 1991).

The most quoted work with respect to the yields of intermediate mass
stars is that of Renzini \& Voli (1981; hereafter RV) who calculated
the amount of matter returned to the ISM by AGB stars in the form of
\eg He, C, N, and O. Their well known results have been widely used to 
compare the predicted abundances in the ejecta of AGB stars with the 
abundances observed in planetary nebulae (see Clegg 1991 and references
therein) and have been often applied in Galactic chemical evolution models
(\eg Matteucci \etal 1989; Rocca-Volmerange \& Schaeffer 1990).

In this paper, we use a synthetic evolution model similar in approach
to that presented by RV to follow the chemical evolution of stars on
the AGB.  However, our model differs substantially from that described
by RV, both in the various aspects of AGB evolution considered as well
as in the parameters that best fit the observations (in particular the
mass loss rate on the AGB).  The model has been described in detail by
Groenewegen \& de Jong (1993; hereafter GJ) and applied to various
observational aspects of AGB evolution, both for AGB stars in the
Galactic disk and Magellanic Clouds (GJ; Groenewegen, van den Hoek \&
de Jong 1995, hereafter GHJ).

An important aspect of AGB evolution largely neglected in previous
studies is the metallicity dependence of the evolutionary algorithms
used.  Observations show that the luminosity function and relative
number ratios of carbon and oxygen-rich AGB stars in the Large and
Small Magellanic Clouds are different (see e.g. GJ). One of the
explanations for this is the different metallicity in these
galaxies. In the actual model, we use a nearly complete metallicity
dependent treatment of the evolution of AGB stars covering the first,
second, and third dredge up.  In addition, in GJ/GHJ take into account 
several new physical
ingredients including the variation of the
luminosity during the interpulse period, the fact that the first few
pulses are not yet at full amplitude, and the detailed inclusion of
mass loss and chemical evolution prior to the AGB.

Before reaching the AGB phase, the main sequence stellar composition
has changed during the first dredge-up (experienced by all stars on
the red giant branch (RGB)) and during the second dredge-up
(experienced by stars with initial mass larger than some certain
critical mass). The first dredge up occurs when the convective
envelope moves inwards as a star becomes a red giant for the first
time so that helium and CNO processed material are brought to the
surface. Several tenths of solar masses can be lost in this phase for
low mass stars (\eg Sweigart et al. 1990; Rood 1973).

The second dredge-up is associated with the formation of the
electron-degenerate CO core after central helium exhaustion and occurs
on the early-AGB (hereafter E-AGB). In this case, helium and nitrogen
may be dredged up towards the stellar surface. We use the
comprehensive set of metallicity dependent stellar evolution tracks
provided by the Geneva group (\eg Schaller et al. 1992) to describe
the evolution prior to the AGB. However, to study in detail the
influence of the first and second dredge up on the AGB yields, we also
consider a metallicity dependent theoretical treatment of these phases
(\cf Sect. 3) according to the recipes outlined in GJ. In both cases,
the stellar evolution prior to the AGB is coupled consistently to the
thermal pulsing AGB phase.

During the third dredge up, carbon is dredged up to the stellar
surface by convection of the carbon-rich pocket formed after each
helium shell flash (or thermal pulse (TP)).  By mixing additional
carbon to the envelope, the star may undergo a transition from M-star
(oxygen-rich), to S-star (carbon roughly equal to oxygen), and C-star
(carbon outnumbering oxygen).  For stars with $m \ga 3-4$ \mss, this
transition is affected by HBB when both carbon already present and
newly dredged-up carbon are processed at the base of the convective
envelope according to the CNO cycle.  We account for the effect of HBB
in an approximate way since the details of this process are not well
understood.  Abundance variations during the AGB of individual
elements heavier than oxygen are not taken into account.

The free parameters in our calculations are the mass loss scaling
parameter $\eta_{\rm AGB}$ for stars on the AGB (using a Reimers law),
the minimum core mass for dredge-up M$_{\rm c}^{\rm min}$, and the
third dredge-up efficiency $\lambda$. We will discuss the effect of
these parameters as well as the effect of HBB on the stellar yields in
Sect. 4. For AGB stars both in the Galactic disk and Magellanic
Clouds, models with $\eta_{\rm AGB}$ = 4, M$_{\rm c}^{\rm min}$ = 0.58
\mss, and $\lambda = 0.75$ including HBB are in best agreement with
the observations. Part of the argumentation for this specific set of
parameter values can be found in GJ and GHJ and references therein.
For this set, we compute the stellar yields of H, $^{4}$He, $^{12}$C,
$^{13}$C, $^{14}$N, and $^{16}$O of AGB stars with initial mass
between 0.8 and 8 \mss, and initial metallicity Z = 0.001, 0.004, 0.008,
0.02 and 0.04, as presented at the end of this paper.

This paper is organized as follows. In Sect. 2, we define the stellar
yields used throughout this paper. In Sect. 3, we briefly describe the basic
ingredients of the synthetic evolution model related to the chemical
evolution of stars on the AGB. In Sect. 4, resulting stellar yields
are presented and their dependence on the main model
assumptions is examined. In Sect. 5, we discuss the preferred set of
stellar yields with respect to earlier theoretical work and compare
the abundances predicted during the final stages of the
AGB with those observed in planetary nebulae (PNe) in the Galactic
disk.

\section{Definition of stellar yields}

The element yield $p_{j}$ of a star of initial mass $m$ is defined as
the newly formed and ejected mass of element $j$ integrated over the
stellar lifetime $\tau(m)$ and normalized to the initial mass (\eg 
Maeder 1992):
\begin{equation}
 mp_{j}(m) = 
\int_{\rm 0}^{\tau(m)} \!\!\! E(m,t) \: . 
\: (Z_{j}(t)-Z_{j}(0))  \:\: {\rm d}t 
\end{equation}
\noindent where $E(m,t)$ denotes the stellar mass-loss rate and
$Z_{j}(t)$ the abundance by mass of element $j$ in the ejecta at
stellar age $t$. Note that negative yields may occur \eg in the case of 
hydrogen consumption.

In general, stellar yields depend on the initial stellar abundances 
in different manners. First, abundances in the 
stellar envelope $Z_{j}(t)$ are related in a complex manner to 
the initial abundances of distinct elements (\eg helium and/or oxygen).
This is particularly true for stars on the AGB as we will discuss below. 
To first order, we take this 
important effect into account by the dependence of the evolutionary algorithms 
used on the initial stellar metallicity $Z(0)$, \ie integrated over 
elements heavier than helium. 
Second, stellar lifetimes $\tau(m)$, remnant 
masses $m_{\rm rem}(m)$, and mass-loss rates $E(m,t)$ vary 
strongly with initial stellar metallicity (\eg Schaller \etal 1992).
Third, stellar yields are defined with respect to the initial stellar
abundances $Z_{j}(0)$ (\cf Eq. 1). 

To allow for a direct comparison of
the derived element yields for pre-AGB and AGB evolution phases (see below),
we adopt the initial abundances as used in the stellar evolution tracks
presented by the Geneva group (\ie Schaller \etal 1992; Schaerer \etal
1993a,b; Charbonnel \etal 1993; Meynet \etal 1994).  
In brief, the Geneva group calculated the initial helium abundance from:
\begin{equation}
Y = Y_{0} + \frac{\Delta Y}{\Delta Z} \; Z
\end{equation}
assuming a primordial helium abundance $Y_{0}$ of 0.24 (\eg Audouze
1987; Steigman 1989) and $\Delta Y / \Delta Z$ = 3 (\eg
Pagel \etal 1986; Pagel \& Kaztauskas 1992) for stars in the Galactic disk.  
Accordingly, 
these tracks imply a revised solar metallicity of $Z_{\odot}=0.0188$ with 
$Y_{\odot}=0.299$ (see Schaller \etal 1992). Initial
abundances of C, N, and O were taken according to the relative
ratios (\cf Anders \& Grevesse 1989) used in the opacity tables by
Rogers \& Iglesias (1992). The hydrogen content was calculated from 
$ X= 1 - Y - Z$.
Table 1 lists the adopted initial abundances of H, $^{4}$He, $^{12}$C, 
$^{13}$C, $^{14}$N, and $^{16}$O at metallicities 
$Z(0)$ = 0.001, 0.004, 0.008, 0.02, and 0.04. Note that abundances are 
given by mass throughout this paper.

\begin{table*}[htp]
\caption[]{Initial element abundances adopted}
\begin{flushleft}
\begin{tabular}{llllll}
\hline
Element  & Z=0.001 & 0.004 & 0.008 & 0.02 & 0.04 \\ \hline
H        & 0.756  & 0.744 & 0.728 & 0.68 & 0.62 \\
$^{4}$He & 0.243  & 0.252 & 0.264 & 0.30 & 0.34  \\
$^{12}$C & 2.24(-4) & 9.73(-4) & 1.79(-3) & 4.47(-3) & 9.73(-3) \\ 
$^{13}$C & 0.04(-4) & 0.16(-4) & 0.29(-4) & 0.72(-4) & 1.56(-4) \\
$^{14}$N & 0.70(-4) & 2.47(-4) & 5.59(-4) & 1.40(-3) & 2.47(-3) \\
$^{16}$O & 5.31(-4) & 2.11(-3) & 4.24(-3) & 1.06(-2) & 2.11(-2) 
\\ \hline
\end{tabular}
\end{flushleft}
\end{table*}

In this paper, we distinguish stellar yields for \eg the pre-AGB and AGB
phases (\cf Sect. 4). In this case, the total mass of
element $j$ ejected during mass-loss phase $i$ (with age boundaries
$t_{\rm l}^{i}(m)$ and $t_{\rm u}^{i}(m)$ in Eq. 1) can be written as:
\begin{equation}
\Delta m_{j}^{i} = \Delta m^{i} Z_{j}(0) + mp_{j}^{i}(m)  
\end{equation}
\noindent where $\Delta m^{i} = \Sigma_{j} \Delta m_{j}^{i}$ is the
total mass ejected during phase $i$.  Similarly, mean abundances of
element $j$ within the ejecta returned to the ISM during phase $i$ can be
written as:
\begin{equation}
<Z_{j}^{i}> = \frac{mp_{j}^{i}}{\Delta m^{i}} + Z_{j}(0)
\end{equation}

The lifetime-integrated stellar yield of element $j$ in terms of the
stellar yields for distinct mass-loss phases $i$ is given by
$mp_{j} = m \Sigma_{i} p_{j}^{i}$. Since
$\Sigma_{j} {\rm p}_{j}$ = 0 and $\Sigma_{j} Z_{j}$ = 1
according to Eq. (1), the total stellar mass ejected can be expressed as: 
$\Delta m^{\rm ej} =
\Sigma_{i} \Delta m^{i} = \Sigma_{i} \Sigma_{j} \Delta m_{j}^{i} = m -
m_{\rm rem}(m)$ where $m_{\rm rem}(m)$ is the stellar remnant
mass. In this manner, Eq. (1) also can be written as:
\begin{equation}
mp_{j}(m) = \Delta m_{j} - \left( m- m_{\rm rem} \right) Z_{j}(0)
\end{equation}

\section{Chemical evolution in the synthetic model}

We briefly repeat that part of the synthetic evolution model that is
related to the chemical evolution on the AGB. Full details on the model can be 
found in GJ. The evolution model is started at the first TP, taking into
account the changes in mass and abundances prior to the first TP, and is
terminated when the envelope mass has been lost due to mass loss or if the 
core reaches the Chandrasekhar mass. The latter situation
never occurs in the best fitting models for the Galaxy and the Large 
Magellanic Cloud (see GHJ, GJ).

\subsection{First dredge-up}

The first dredge-up occurs when the convective envelope moves inwards
as a star becomes a red giant for the first time. 
Description of the mass loss during the RGB can be found in GJ.
The convective motion dredges up material that was previously located near the
hydrogen burning shell. The increase in the helium abundance, $\Delta Y$, 
is given by (\cf Sweigart \etal 1990):
\begin{equation}
\hspace{-0.5cm} \Delta Y = \left\{ \begin{array}{ll}
      -0.0170 \; m + 0.0425 & \mbox{for $m < 2, \; Y = 0.3$}  \\
      -0.0068 \; m + 0.0221 & \mbox{for $2 \leq m < 3.25, \; Y = 0.3$}  \\
      -0.0220 \; m + 0.0605 & \mbox{for $m < 2.2, \; Y = 0.2$}  \\
      -0.0078 \; m + 0.0293 & \mbox{for $2.2 \leq m < 3.75 , \; Y = 0.2$}  \\
       0    & \mbox{else}
\end{array}
\right.
\end{equation}
Results are linearly interpolated for a given $Y$ while the small dependence of 
$\Delta$Y on Z for a given Y is neglected.
The change in  hydrogen is opposite to the change in helium:
\begin{equation}
\Delta X = - \Delta Y
\end{equation}
Changes in $^{12}$C, $^{14}$N, and $^{16}$O are calculated from:
\begin{equation}
\begin{array}{llll}
\Delta ^{12}C & = & ^{12}C \; (g-1)     &       \\
\Delta ^{14}N & = & -1.167 \; \Delta ^{12}C   \\ 
\Delta ^{16}O & = & -0.01 \; ^{16}O     &     \\
\end{array}
\end{equation}
\begin{equation}
g = \left\{ 
\begin{array}{ll}
           0.64 - 0.05 \; (m - 3) & \mbox{for $m < 3$} \\
           0.64                   & \mbox{for $m \geq 3$} \\
\end{array}
\right.
\end{equation}
\noindent The number ratio $^{12}C$/$^{13}C$ after the first dredge-up
does not vary much with mass or composition (Sweigart \etal 1989) and
is set to 23.

\subsection{Second dredge-up}

The second dredge-up is related to the formation of the
electron-degenerate CO core after central helium exhaustion
in stars more massive than a critical mass (see Becker \& 
Iben 1979, 1980) which depends both on the core mass and main-sequence 
abundances. The base of the convective envelope
moves inward through matter pushed outwards by the He-burning shell.
The treatment of the second dredge-up follows that of RV closely.
No mass loss is assumed during second dredge-up.

Following GJ (and references therein), the abundances after the second
dredge-up can be obtained from the abundances prior to the second
dredge-up and the abundances of the material that is dredged up using
the relation:
\begin{equation}
 X^{\rm after} = a X^{\rm prior} + b X^{\rm du}
\end{equation}
where the coefficients $a$ and $b$ are functions of the total mass as well 
as the core
mass before and after second dredge-up and are given by Eq. (30) in GJ.
The abundances prior to the second dredge-up are known and the
abundances of the dredged up material are given by RV and Iben \& Truran 
(1978):
\begin{eqnarray}
Y^{\rm du} & = & 1 - Z  \nonumber \\
^{14}N^{\rm du} & = & 14 \; (^{12}C/12 + ^{13}C/13 + ^{14}N/14 + ^{16}O/16) 
      \nonumber \\
^{12}C^{\rm du} & = & ^{13}C^{\rm du} =  ^{16}O^{\rm du} = 0
\end{eqnarray}
The hydrogen abundance after second dredge-up was
calculated from $X = 1 - Y - Z$, with $Y$ and $Z$ the helium and metal
abundance after second dredge-up. This was done to ensure that $X + Y +
Z \equiv 1$ at all times.

\subsection{Third dredge-up}

As discussed above, the pre-AGB evolution with respect to mass loss
and chemical evolution during the first and second dredge-up is either
calculated according to the recipes in GJ (see previous subsections),
or is taken from the theoretical evolution tracks provided by the
Geneva group. In the latter case, we use the stellar surface
abundances as well as the stellar mass at the end of these tracks. 

The synthetic AGB evolution model starts at the first thermal pulse.
In brief, we account for the dependence of core mass on initial
stellar metallicity and assume that third dredge-up occurs only if the
core mass is larger than a critical value $M_{\rm c}^{\rm min}$.  In
GJ we argued that a value of $M_{\rm c}^{\rm min} \sim$ 0.58 \ms is
required to fit the low-luminosity tail of the observed carbon star
luminosity function in the LMC (see below).

The time scale on which thermal pulses occur is a function of core
mass as discovered by Paczynski (1975). In GJ and GHJ, we use the
core-mass-interpulse relation presented in Boothroyd \& Sackmann
(1988) where the increase in core mass during the interpulse period
($t_{\rm ip}$) is given by:
\begin{equation}
\Delta M_{\rm c} = \int_{o}^{t_{\rm ip}} \frac{d M_{\rm c}}{{\rm d} t}
{\rm d}t 
\end{equation}
A certain fraction of this amount is assumed to be dredged up:
\begin{equation}
\Delta M_{\rm dredge} = \lambda \, \Delta M_{\rm c}
\end{equation}
The free dredge-up parameter $\lambda$ is assumed to be a constant.
In GJ we found that a value of $\lambda$ = 0.75 is required to fit the
peak of the observed carbon stars LF in the LMC (see below).

In principle, the composition of the dredged-up material is determined
by the detailed chemical evolution of the core. For simplicity, we
assume that the composition of the material dredged-up after a TP is:
$^{4}$He = 0.76, $^{12}$C = 0.22, and $^{16}$O = 0.02 (\cf Boothroyd
\& Sackmann 1988).  The carbon is formed through incomplete helium
burning in the triple $\alpha$ process and the oxygen through the
$^{12}$C($\alpha, \gamma)^{16}$O reaction.

Newly dredged-up material can be processed at the base of the
convective envelope in the CNO-cycle, a process referred to as hot
bottom burning (HBB) and extensively discussed by RV. To a large
extent, HBB determines the composition of the material in the stellar
envelope of thermal pulsing AGB stars.  The process of HBB is able to
slow down or even prevent the formation of carbon stars (\eg Groenewegen \& 
de Jong 1994a). Since
$^{12}$C is converted into $^{13}$C and $^{14}$N, it also gives rise
to the formation of $^{13}$C-rich carbon stars (usually referred to as
J-type carbon stars) and $^{14}$N-rich objects (\eg Richer \etal
1979).

RV treated HBB in considerable detail as a function of the mixing
length parameter (\eg $\alpha$ = 0, 1.0, 1.5, 2).  In GJ (see for
details their Appendix A) it was decided to approximate in a
semi-analytical way the results of RV for their $\alpha$ = 2 case as
it gave the largest effect of HBB.  Since then new results regarding
HBB have been obtained, both theoretically (Boothroyd et al. 1993,
1995) and observationally for AGB stars in the Magellanic Clouds (Plez
et al. 1993; Smith et al. 1995).  These results suggest that HBB is a
common phenomenon that occurs at a level roughly consistent with that
predicted by RV in case $\alpha$ = 2.  In particular, Boothroyd et
al. (1995) estimate that the initial stellar mass above which HBB
takes place is $\sim$4.5 \ms which is similar to the value of
$\sim$3.3 \ms predicted by RV ($\alpha$=2).  Observations indicate
that virtually all stars brighter than $M_{\rm bol} \approx -6$ mag
undergo envelope burning (Smith et al. 1995). These luminosities are
reached for stars with initial masses slightly below 4 \ms and larger
(Boothroyd et al. 1993).

During the thermal pulsing AGB, stars lose most of their mass: typically
$\sim$ 0.4 \ms for a 1 \ms star and $\sim $4.8 \ms for a 6 \ms star (at solar 
initial metallicity; see GJ). Clearly, stellar yields for intermediate 
mass stars are dominated by the mass loss and chemical evolution during this 
phase. After gradual ejection of their outer envelope, most AGB stars leave a 
white dwarf remnant usually accompanied by the formation of a planetary nebula 
(PN).

\subsection{Tuning, calibration, and uncertainties of the synthetic 
evolution model}

We briefly discuss assumptions and uncertainties involved with the 
synthetic evolution model described above.
In GJ and GHJ we used observations of AGB stars in the LMC and Galaxy
to constrain the synthetic AGB model. The main model parameters are the
minimum core mass for third dredge-up $M_{\rm c}^{\rm min}$, the third
dredge-up efficiency $\lambda$, and the scaling law for mass loss on
the AGB ${\eta}_{\rm AGB}$.  

In GJ the following observational constraints were considered: the
luminosity function of carbon stars, the observed number ratio of
carbon to oxygen-rich AGB stars, the birth rate of AGB stars, the
abundances observed in PNe, the initial-final mass relation, and the
frequency of carbon stars in clusters of a given mass. Values of
$M_{\rm c}^{\rm min}$ = 0.58 \ms and $\lambda$ = 0.75 were determined
predominantly by fitting the carbon star LF.  A value of ${\eta}_{\rm
AGB} = 4$ (assuming a Reimers mass loss law) was derived by fitting
the high-luminosity tail of this LF.  This set of model parameters
resulted in a birthrate of AGB stars consistent with other
determinations and predicts that carbon stars can form only from
main-sequence stars more massive than 1.2 \mss, consistent with
observations of carbon stars in LMC clusters.  In Groenewegen \& de
Jong (1994a) we showed that the model with the aforementioned
parameters predicted the correct observed abundances in LMC PNe. In
Groenewegen \& de Jong (1994b) we considered two alternative mass loss
formula to the Reimers one. We found that the mass loss formula
proposed by Vassiliadis \& Wood (1993) could less well explain the
observed abundances while a scaled version of the law proposed by
Bl\"ocker \& Sch\"onberner (1993) could equally well fit the
observations.

In GHJ we applied the synthetic evolution model to carbon stars in the
Galactic disk. As there yet exists no reliable carbon star LF, we used
observations of carbon stars in open clusters and in binaries to
determine $M_{\rm c}^{\rm min}$ = 0.58 \mss.  Using similar
constraints as for AGB stars in the LMC, values of $\lambda = 0.75$
and ${\eta}_{\rm AGB} = 4$ were found in optimal agreement with the
observations.  Thus, models for the observed luminosity function of
carbon stars in both the Galactic disk and LMC do favour a high mass
loss coefficient $\eta_{\rm AGB} \approx 3-5$.  This range in
$\eta_{\rm AGB}$ is consistent with additional constraints such as the
initial-final mass relation for C$+$O white dwarfs in the solar
neighbourhood (see Weidemann \& Koester 1983; Weidemann 1990). 
In conclusion, a wide range of
observations of AGB stars both in the LMC and Galactic disk can be
explained by one and the same set of model parameters, \ie $M_{\rm
c}^{\rm min}$ = 0.58 \mss, $\lambda =0.75$, and ${\eta}_{\rm AGB} =
4$. In the following, we will refer to this set of parameters as the
standard model.

It should be noted that observations indicate that some stars do not
obey the standard model predictions. In particular, the
$^{12}$C/$^{13}$C ratio after the first dredge-up is often lower than
predicted in stars of low mass, down to $^{12}$C/$^{13}$C $\approx$
10. Rotationally induced mixing (\eg Sweigart \& Mengel 1979; Charbonnel 1995; 
Denissenkov \& Weiss 1996) or initial abundances different from
those adopted here (see Sect. 4) may play a role.

Clearly, the assumptions of a fixed critical core mass as well as of a
constant dredge-up efficiency and mass loss parameter for all AGB
stars (independent of their initial mass and AGB phase) are first
order approximations.  There is much debate whether or not material is
dredged up at every thermal pulse, and how much.  Furthermore, it
seems possible that the dredge-up process is turned off when a star
becomes a carbon star (\eg Lattanzio 1989).  Notwithstanding, this
simple three parameter model can explain essentially all present-day
observations of AGB stars so there appears no need for a more
complicated model (although this does not prove that our model
assumptions are correct). We will investigate the sensitivity of the
stellar yields on the adopted values of M$_{\rm c}^{\rm min}$,
$\lambda$, and $\eta_{\rm AGB}$ in Sect. 4.

An additional source of uncertainty is associated with the number of
atoms (and isotopes) taken into account, \ie H, He, $^{12}$C,
$^{13}$C, $^{14}$N and $^{16}$O.  The first and second dredge-up
abundance changes are either taken directly from the model tracks of
the Geneva group, or, in the synthetic model, through parametrisation
of other model calculations. All these works include a much larger
chemical network than that considered here.  The third dredge-up is
simplified in the sense that only the $^{12}$C($\alpha, \gamma)^{16}$O
is included and that the abundances of $^{12}$C, $^{16}$O and He in
the convective zone after a TP are taken from Boothroyd \& Sackmann
(1988) and are assumed to be constant. In particular, we do not
consider s-process reactions which take place in the convective
inter-shell. The most important effect of this process on the species
we consider is through the $^{13}$C($\alpha, {\rm n})^{16}$O
reaction. However, the amount of matter burnt in this reaction is
probably small (Marigo \etal 1996) and apparently depends on the
amount of semi-convection assumed in the models (Busso et al. 1992,
1995).

Other uncertainties concern the detailed inclusion of HBB. In
particular, the temperature structure of the envelope, the fraction of
dredged up material processed in the CNO cycle, and the amount of
envelope matter mixed down and processed at the bottom of the envelope
may vary among AGB stars differing in initial mass, composition, and
age. Nevertheless, although the details on HBB are poorly understood
yet, good agreement is obtained between the standard model predictions
including HBB and observations related to HBB in massive AGB stars.
We will investigate the effect of HBB on the stellar yields by
introducing the parameter $m_{\rm HBB}$, which defines the core mass
at which HBB is assumed to operate (according to the recipes outlined
in the Appendix in GJ). The default value used in the standard model
is $m_{\rm HBB}$ = 0.8 \ms which de facto is the value used in GJ and
GHJ.  Other values of $m_{\rm HBB}$ are discussed below.

\section{Results}

In Tables 2-21, we present theoretical yields p$_{j}$ for AGB stars in the
mass range 0.8$-$8 \ms in case of the standard model ($\eta_{\rm AGB}$
= 4, $M_{\rm c}^{\rm min}$ = 0.58 \mss, $\lambda$ = 0.75, and $m_{\rm
HBB}$ = 0.8 \mss) including first, second, and third dredge-up as well
as HBB.  We distinguish pre-AGB, AGB, final AGB, and total element
yields at initial metallicities Z = 0.001, 0.004, 0.008, 0.02, and 0.04.
Pre-AGB yields are the yields up to the end of the E-AGB. AGB yields
are the yields on the TP-AGB {\em except} for the last 2.5 10$^{4}$
yr. The final AGB yields are the yields on the TP-AGB integrated over
the last 2.5 10$^{4}$ yr. This distinction is made to compare with
the abundances in PNe (note that final AGB yields have been omitted if 
the AGB lifetime is much smaller than 2.5 10$^{4}$ yr).

Pre-AGB evolution is based on the Geneva group (\eg Schaller \etal
1992). These uniform grids of stellar
models are based on up-to-date physical input (\eg opacities, nuclear
reaction rates, mixing schemes, etc.) and cover the relevant initial 
stellar mass range from 0.8 to 8 \ms as well as initial metallicity from
Z = 0.001$-$0.04. For stars with $m \la$ 1.7 \ms these tracks have been 
computed up to the He flash, for 2 $<$ $m$ $<$5 \ms up to the E-AGB, and for 
$m > 7$ \ms until the end of central C-burning. Recently, Charbonnel et al. 
(1996) presented new grids of models covering the evolution (from the zero 
age main sequence up to the end of the E-AGB)
of low mass stars with initial masses between 0.8 and 1.7 \ms born with 
metallicities $Z$ = 0.02 and 0.001. We note that these stars were not
evolved through the helium core flash but instead were evolved from 
constructed zero-age horizontal branch models (see Charbonnel et al. 1996).

For stars with initial mass above 1.25 \mss, the Geneva tracks
used are with overshooting and standard mass loss rates (see \eg
Schaller \etal). For stars below 1.25 \mss, the tracks used are
without overshooting (for $m = 1.25$ \ms we include yields both for
tracks with and without overshooting).  

We neglect the fact that the Geneva tracks for stars with $m\la 1.7$
\ms and $Z$ = 0.004, 0.008, and 0.04 end at the helium flash and do
not extent to the end of the E-AGB. However, these low mass stars do
not experience the second dredge-up and are expected not to loose much
mass on the horizontal branch and E-AGB, so that the influence on the
yields is negligible (see Sect. 4.4).

In Tables 2$-$21, we list subsequently the initial mass $m_{\rm ini}$,
element yields p$_{j}$ of H, $^{4}$He, $^{12}$C, $^{13}$C, $^{14}$N, 
$^{16}$O,
integrated CNO-yield Y$_{\rm CNO}$, total element yield Y$_{\rm tot}$
(elements heavier than helium), the total amount of mass returned
$\Delta m_{\rm ej}$, and the stellar mass $m_{\rm end}$ at the end of
each phase.

We consider resulting stellar yields for various choices 
of the Reimers mass loss coefficient $\eta_{\rm AGB} =$ 1$-$5, third
dredge-up efficiency $\lambda =$ 0.6$-$0.9, critical core mass for
dredge up $M_{\rm c}^{\rm min} = $ 0.56$-$0.62 \mss, and minimum core
mass for HBB $m_{\rm HBB}$ = 0.8 and 0.9 \ms (see Sect. 3). We examine
the impact of these quantities as well as of the adopted pre-AGB
evolution model on the predicted yields.

\subsection{Dependence on mass loss}

Figure 1 shows resulting AGB yields for various values of the Reimers
mass-loss parameter $\eta_{\rm AGB}$ = 1$-$5. The other parameters are
as for the standard model (unless stated otherwise).  Low
mass AGB stars ($m \la$ 4 \mss) predominantly contribute to helium and
carbon. High mass AGB stars are important contributors to helium and
nitrogen (see below). For the standard model, element yields are
smaller by factors typically 2$-$3 compared to the $\eta_{\rm AGB}
\sim 1$ case.  Resulting yields increase with decreasing values of
$\eta_{\rm AGB}$ (\ie smaller mass-loss rates) as a lower value of
$\eta$ results in longer AGB lifetimes and therefore more thermal
pulses (assuming that the amount of dredged-up matter during a thermal
pulse is roughly constant). For large values of $\eta_{\rm AGB} \ga
5$, the effect of increasing $\eta_{\rm AGB}$ on both the AGB lifetimes and
number of thermal pulses becomes negligible and the predicted yields
remain approximately constant.

\begin{figure*}[htp]
\leftline{\psfig{figure=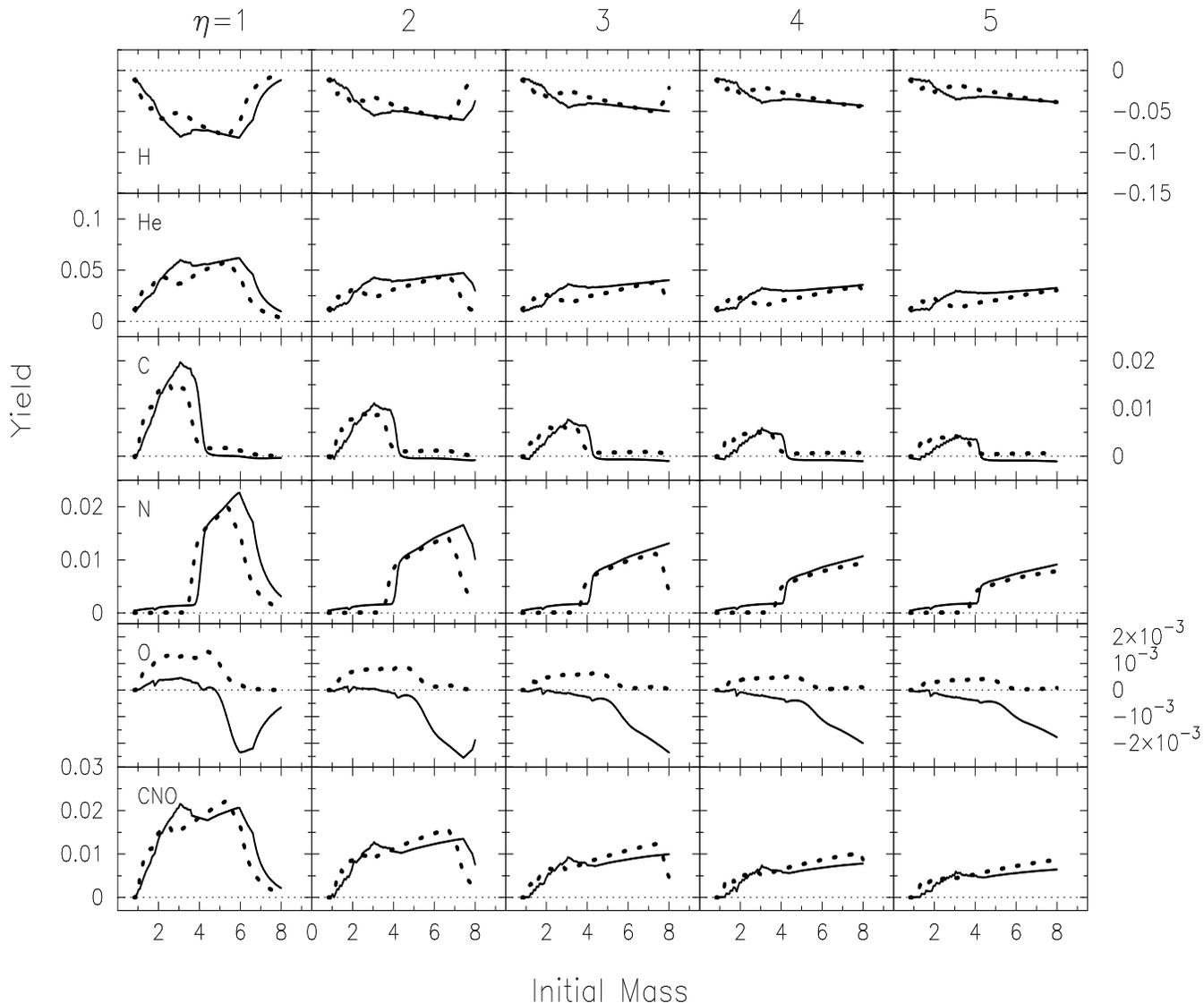,height=16.cm,width=18.cm,angle=270.}}
\caption[]{Stellar yields of H, $^{4}$He, $^{12}$C, $^{14}$N, $^{16}$O, and 
total CNO vs. initial stellar mass for $\eta_{\rm AGB}$=1$-$5 and initial 
compositions (Z, Y) = (0.02, 0.32; solid line) and (0.001, 0.24; dotted).
Parameters values are further as for the standard model (\ie 
$\eta_{\rm AGB}$ = 4)}
\end{figure*}

\subsection{Dependence on initial metallicity}

In general, carbon and oxygen yields increase with decreasing initial
metallicity $Z$ (\cf Fig. 1). This is due to the fact that dredge-up with
subsequent CNO-burning affects more strongly the composition of
envelopes with relatively low initial abundances (see \eg RV). In
addition, the core mass at the first thermal pulse is larger at low
metallicities (GJ).  Therefore, the amount of material dredged-up from
the core to the envelope is substantially larger in initially low$-Z$
AGB stars. In contrast, nitrogen yields slightly increase with
metallicity as nitrogen is formed during CNO-burning by consumption of
C and O. For hydrogen and helium, the sensitivity of the yields to 
initial metallicity are mainly due to the effect of dredge-up, \ie post 
dredge-up processing of H and He is usually low. Note that even 
small changes in the yields of AGB stars due to variations in initial 
metallicity can significantly affect the enrichment of the ISM (after
weighting by the initial mass function and star formation rate at the 
time these stars were formed).

We like to emphasize that AGB yields of intermediate mass stars 
are strongly dependent on the abundances of distinct elements (\eg C, N,
and O) in the galactic ISM from which these stars are formed. In other 
words, stars with initial element abundances substantially different 
from those listed in Table 1 have AGB yields distinct from the yields given 
in Tables 2-21. We will return to this important point below when model 
predictions are confronted with abundances observed in PNe in the Galactic 
disk.

\subsection{Dependence on the amount of HBB}

As discussed before, HBB may prevent or slow down the formation of 
carbon stars by the possible destruction of newly dredged up carbon at the 
base of the convective envelope. Fig. 2 illustrates that 
for low mass AGB stars ($m$ $<$ 4 \mss), the effect 
of HBB is negligible due to the low temperature at the bottom of 
their envelopes (GJ). For high mass AGB stars, the effect of HBB 
depends on the amount of matter exposed to the high temperatures at the 
bottom of their envelopes, the net result being the conversion of carbon and 
oxygen to nitrogen. Yields of H, He, and total CNO are not affected by HBB 
since basically two reaction chains of the CNO-cycle are involved, \ie 
the CN and ON-cycle.

\begin{figure*}[thp]
\leftline{\psfig{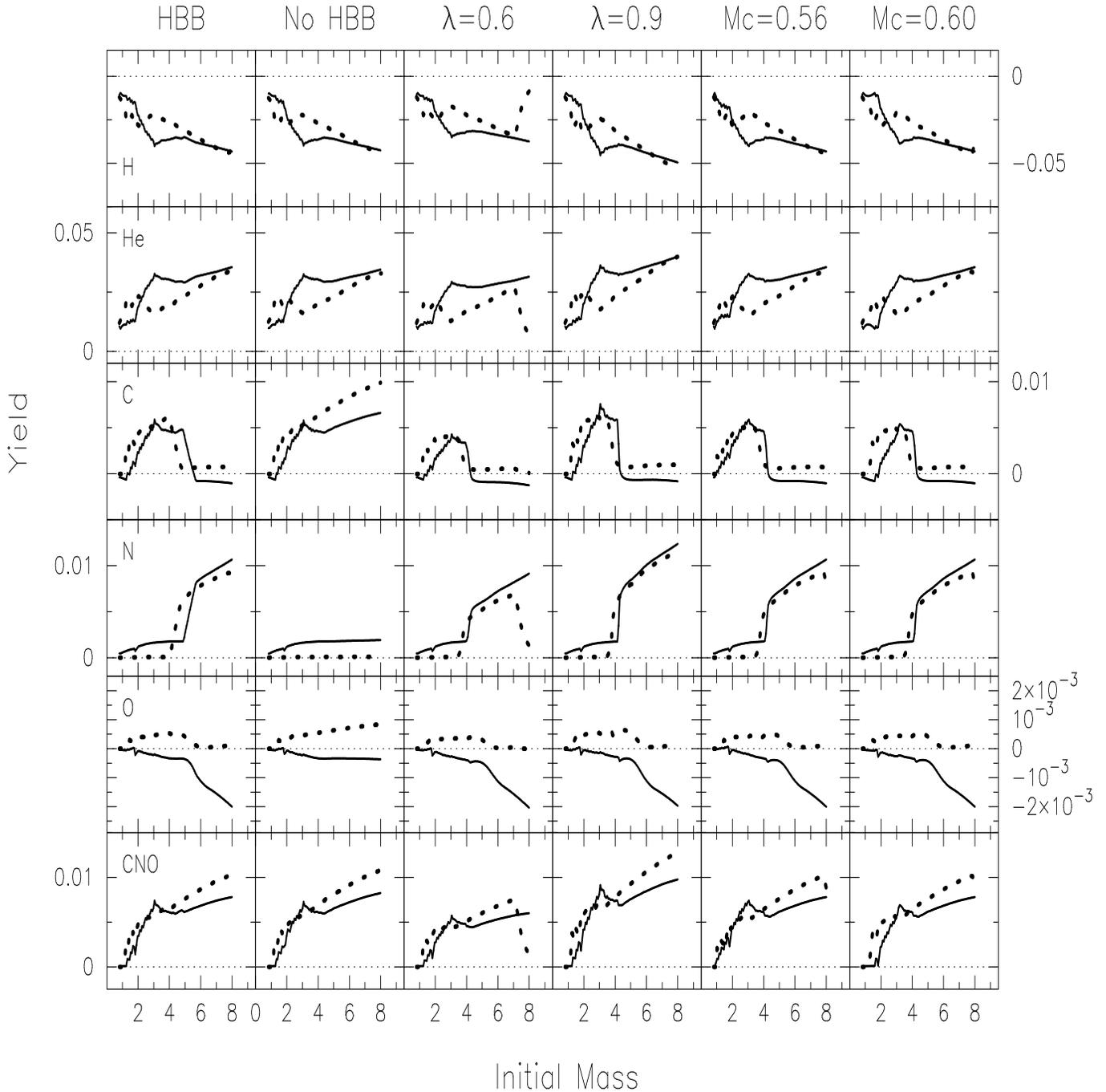}}
\caption[]{Stellar yields of H, $^{4}$He, $^{12}$C, $^{14}$N, $^{16}$O, and 
total CNO vs. initial stellar mass for the standard model: effect of 
varying 1) the amount of HBB ({\em first two columns}), 2) the 
dredge-up efficiency ({\em center columns}), and 3) the critical core mass 
for dredge-up ({\em last two columns})}
\end{figure*}

We compare in Fig. 2 the resulting yields in case $m_{\rm HBB}$ = 0.9 
(figure labelled HBB)
and 1.3 \ms (no HBB), respectively.  A choice of $m_{\rm HBB}\approx $ 1.3 \ms
or larger results in no HBB as none of the AGB stars in our model
reach such high core masses.  In case of HBB, the strong decrease of
the carbon and strong increase in the nitrogen yield can be seen at
masses at $m \ga 4-5$ \mss.  In the no HBB case, the stellar yields of
carbon are seen to dominate the total CNO-yields over the entire mass
range. 

The effect of changing $m_{\rm HBB}$ from 0.8 to 0.9 \ms is that HBB
operates in stars of initial mass $\ga 5$ \ms instead of $\ga 4$
\mss. Since $m_{\rm HBB}$ = 0.9 \ms provides a reasonable upper limit to 
the minimum stellar mass experiencing HBB, we have included in 
Tables 22-31 $^{12}$C, $^{13}$C, $^{14}$N, and $^{16}$O yields for stars 
more massive than 4 \ms in case of the standard model with $m_{\rm HBB}$ = 
0.9 \ms (pre-AGB are as given at the corresponding metallicities in 
Tables 2-21). These yields include the minimum effect of HBB as 
inferred from the observations and are somewhat smaller than those given
by the standard model (\ie $m_{\rm HBB}$ = 0.8 \mss).

As discussed earlier, the default choice of $m_{\rm HBB}$ = 0.8 \ms
is based on our implementation of the RV $\alpha$ = 2 model but
appears consistent with recent observations of HBB in AGB stars 
both in the SMC and LMC as well as recent model calculations on massive 
AGB stars. In any case, HBB is required to explain observations. More 
observations are needed to investigate any dependence of HBB on metallicity.

\subsection{Dependence on third dredge up efficiency}

We consider in Fig. 2 AGB yields for third dredge-up efficiencies $\lambda =
0.6$ and 0.9 (\ie the range allowed for by the observations; GJ). Stellar 
yields increase substantially when
$\lambda$ is increased, \ie enhancing the amount of carbon and helium
that is dredged-up and added to the stellar envelope after each
thermal pulse. In addition, the composition of dredged-up material may be 
strongly affected by HBB, in particular for high mass stars. In other words,
increasing $\lambda$ leads to an increase in the carbon yields for low
mass stars and to an increase in nitrogen for high mass stars. Furthermore, 
helium yields increase for all stars with initial masses above
$\approx$ 1.5 \ms which corresponds to the limit of $M_{\rm c}^{\rm min}$ 
= 0.58 \mss. 

\subsection{Dependence on critical core mass for dredge up}

Yields for extreme values of the minimal core mass for third dredge-up  
$M_{\rm c}^{\rm min}$ = 0.56 and 0.60 \mss, respectively, are shown in 
last two columns of Fig. 2. The effects of varying $M_{\rm c}^{\rm min}$ 
are limited to relatively low mass AGB stars ($\la$ 2 \mss). 
A larger value of $M_{\rm c}^{\rm min}$ implies a higher initial mass for 
stars that can turn into carbon stars. This results in negative carbon yields
(corresponding to the depletion of carbon during first dredge-up) over
a larger range in initial mass (helium yields decrease over this mass 
range as well). A value of $M_{\rm c}^{\rm min}$ as small as $\sim 0.56$ \ms
would imply that all AGB stars end as carbon stars while $M_{\rm c}^{\rm min}
\ga 0.61$ \ms would inhibit carbon star formation. Clearly, the third 
dredge-up and the precise values of $M_{\rm c}^{\rm min}$ and $\lambda$ 
are of crucial importance for the formation of carbon stars.
We like to emphasize that the parameter value ranges consistent with the 
observations are rather narrow and are mutually correlated
(\eg in case of $M_{\rm c}^{\rm min}$ and $\lambda$).

\subsection{Dependence on pre-AGB evolution}

In GJ and GHJ the description of the pre-thermal pulsing AGB evolution
was taken from recipes in the literature or fits made to published
results.  An alternative approach is to directly use stellar evolution
tracks, as is done in this study (see above).  

In Fig. 3, we compare for $Z$ = 0.02 and 0.001 the resulting AGB
yields in the case of pre-AGB evolution according to the evolution
tracks described above with those computed following the recipes from
GJ/GHJ. In both cases, the initial stellar composition has been
adopted from the Geneva group to comply with the stellar evolution
tracks prior to the AGB.

\begin{figure*}[thp]
\leftline{\psfig{figure=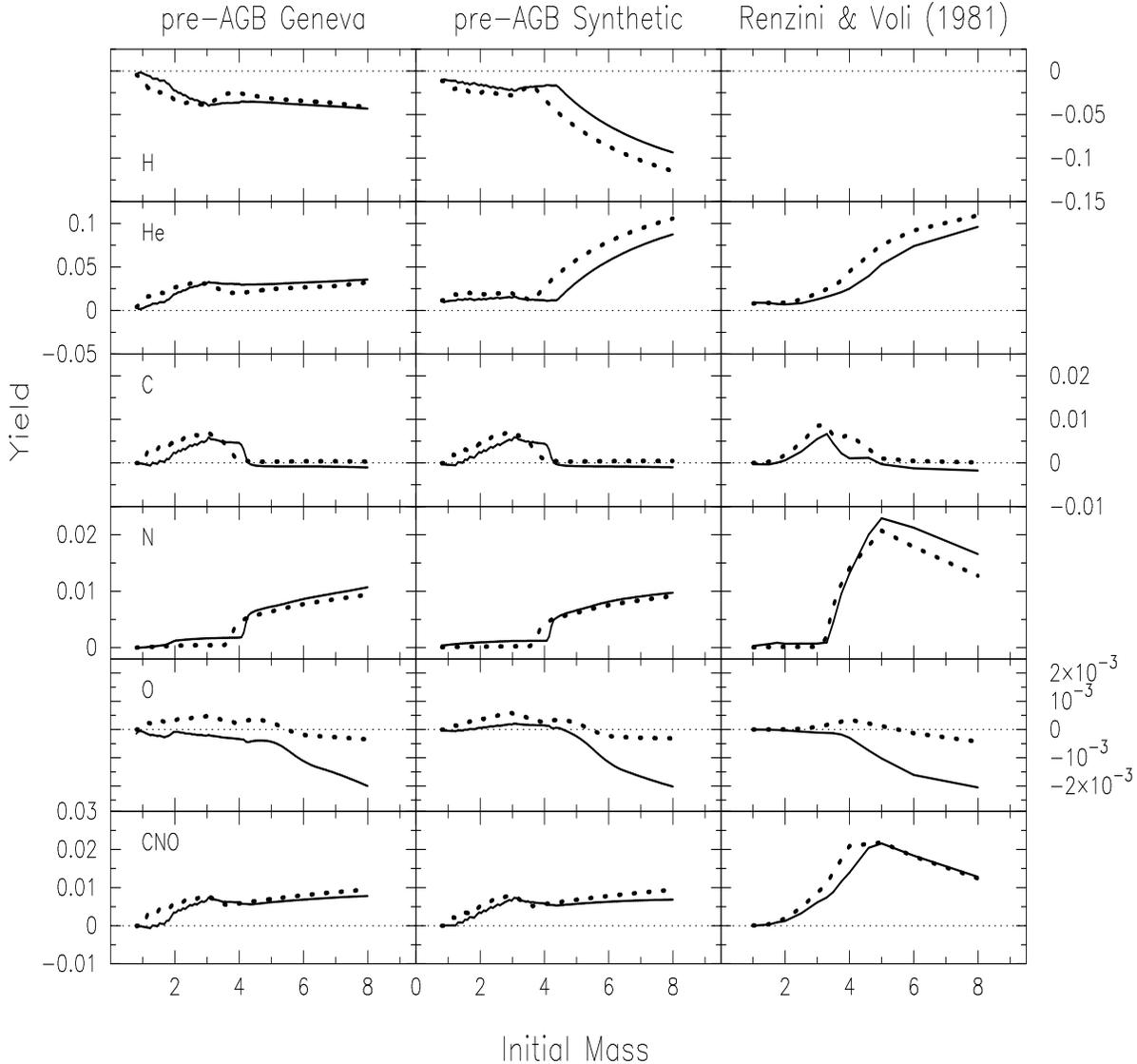,height=16.cm,width=16.cm,angle=270.}}
\caption[]{Stellar yields including HBB vs. initial stellar mass of H, 
$^{4}$He, $^{12}$C, $^{14}$N, $^{16}$O, and total CNO in case of the standard 
model for different treatments of pre-AGB evolution (left and center panels)
and initial metallicities $Z$ = 0.02 (solid line) and 0.004 (dotted line; 
yields for $Z$=0.001 are similar). 
The yields presented by RV (their case $\alpha = 2$)
at initial compositions $(Z, Y)$ = (0.02,0.32; solid lines) and (0.004,0.24; 
dotted) are shown for comparison (hydrogen yields are not given explicitly 
by RV)} 
\end{figure*}

Differences between the two sets of yields are found to be very small
except for hydrogen and helium where the GJ/GHJ approach predicts
higher yields for massive stars. This is traced back to differences in
the treatment of the second dredge-up process. We list in Table 32 the
corresponding total AGB yields of H and He for the synthetic GJ/GHJ
model (initial metallicities as before).  These modified yields can be
used when the effect of 2nd dredge up is expected to be more
pronounced than that given by the Geneva group (see below). The larger
yields also imply higher helium abundances which has interesting
consequences for the predicted helium abundances in planetary nebulae
(see Sect. 5).

We find that pre-AGB evolution for stars born with metallicity $Z$ =
0.02 is in general unimportant for the total yields of AGB stars,
except in case of helium and hydrogen.  This confirms our earlier
statement that the thermal pulsing AGB is the most important phase in
determining the yields of intermediate mass stars (even though initial
stellar abundances, effects of overshooting, mixing, and pre-AGB
evolution can play a significant role).

\subsection{Conclusion}

We conclude that the resulting AGB yields are most sensitive to the
mass loss parameter $\eta_{\rm AGB}$, the effect of HBB, and the
initial stellar abundances. Variations in the remaining model
parameters result in stellar yields not substantially different from
those predicted by the standard model with $\eta_{\rm AGB}$ = 4,
$M_{\rm c}^{\rm min} = 0.58$ \mss, $\lambda=0.75$, and $m_{\rm HBB}$ =
0.8 \mss. Stellar yields for model parameters distinct from that used
for the standard model (\eg $\eta_{\rm AGB}$, HBB, or initial
abundances) are available upon request.  We like to emphasize that small 
differences in the predicted AGB yields may be important when integrating 
over the initial mass function in galactic chemical evolution models.

\section{Discussion}

\subsection{Comparison with RV}

We compare in Fig. 3 the yields predicted by the standard model with those 
given by RV for stars formed with initial metallicities 
$Z$ = 0.02 ($Y$ = 0.28) and $Z$ = 0.004 ($Y$ = 0.232), respectively.
We take into account the observational fact that HBB takes place in stars 
more massive than $\sim$3.5 \mss. Accordingly, we use the yields given by RV
in case $Z$ = 0.02 with $\alpha$ = 2.0 for $m > 3.3$ \ms and $\alpha = 0.0$ 
for $m \la 3$ \ms (\ie their tables 3e and 3a, respectively). Similarly,
we use their yields in case $Z$ = 0.004 with $\alpha = 1.5$ for $m > 3.25$ \ms 
and $\alpha=0.0$ for $m < 3$ \ms (\ie their tables 3i and 3h). For 
$Z$=0.004 RV did not tabulate results for $\alpha = 2$.

Fig. 3 shows that the standard model (pre-AGB evolution as described 
in Sect. 3) results in yields which differ from the selected yields of RV
within a factor 2$-$3. The standard model predicts larger yields 
predominantly for AGB stars with $m \la 4$ \ms in case of carbon and oxygen, 
and for AGB stars more massive than $\sim$3.25 \ms in case of nitrogen.
In this comparison, we have neglected the effect on the resulting yields
of differences (up to $\sim$ 25 \%) in the initial C, N, and O abundances 
between the standard model and that of RV. 

It was derived in GJ that $\eta_{\rm AGB} \ga 3$ is needed to fit the
observed initial-final mass relation for stars in the Galactic disk
(see Weidemann \& Koester 1983).  However, the selected yields of RV
were computed for $\eta_{\rm AGB}$ = 0.33 while for the standard model
$\eta_{\rm AGB}$ = 4. Therefore, due to the high mass loss rates of AGB
stars much fewer thermal pulses on the AGB occur in the standard model
compared to the RV models.  The main part of the observed differences
between the yields predicted by the standard model and those of RV are
probably due to this effect (apart from differences in the detailed
description of evolution along the AGB, in particular the efficiency of third dredge-up).

In conclusion, we find that the selected yields given by RV differ
from those predicted by the standard model by a factor 2$-$3.  In
particular, for high mass AGB stars ($m \ga 3.5$ \mss), the effect of
HBB on the nitrogen yields for the selected RV models is much larger
than that for the standard model. This suggests that values of the
mixing length parameter $\alpha <$ 2 may be more appropriate for
massive AGB stars as we will argue below.

\subsection{AGB stars and the enrichment of the galactic ISM}

We have presented the yields of intermediate mass AGB stars for
appropriate ranges in mass, initial composition, mass loss parameter
$\eta_{\rm AGB}$, and effects of second dredge-up and HBB.  We have
shown that the yields of such stars are determined by their final
stages and are important for the carbon and nitrogen enrichment of the
Galactic disk ISM. In particular, AGB stars account probably for more
than 90 \% of the interstellar nitrogen in the disk (depending on the
shape of the IMF at low and intermediate mass stars).

From the results presented in GJ and GHJ, we argued that the standard
model with ${\eta}_{\rm AGB} \sim 4$ provides a reasonable
approximation of the yields of intermediate mass AGB stars in the
Galactic disk and the Large Magellanic Cloud. These systems have a
metallicity that differ by only a factor of 2 (\eg Russell \& Dopita
1992). In galaxies with a substantial lower metallicity, one may
expect a lower value of ${\eta}_{\rm AGB}$ to be more appropriate. We
like to emphasize that using a fixed value of ${\eta}_{\rm AGB}$ does
not necessarily mean identical mass loss rates as two stars of the
same initial mass evolve differently in the synthetic model due to the
explicit metallicity dependence of the recipes used.

Direct observational information on the metallicity dependence of mass
loss and element yields in AGB stars is rare. In Groenewegen et
al. (1995), the spectral energy distributions and 8-13 $\mu$m spectra
of three long-period variables (one each in the SMC, LMC and Galaxy)
with roughly the same period were fitted. From the derived ratios of
the dust optical depths in these stars, it was argued that the mass
loss rates of AGB stars in the Galaxy, LMC, and SMC are roughly in the
ratio of 4:3:1. This corroborates that ${\eta}_{\rm AGB}$ could be
similar for AGB stars in the Galaxy and LMC. Furthermore, this
suggests that for AGB stars in low metallicity systems like the SMC
($Z$ $\le$ 0.004), values of ${\eta}_{\rm AGB}$ $\approx$ 1$-$2 may
be more appropriate. 

\begin{figure*}[htp]
\leftline{\psfig{figure=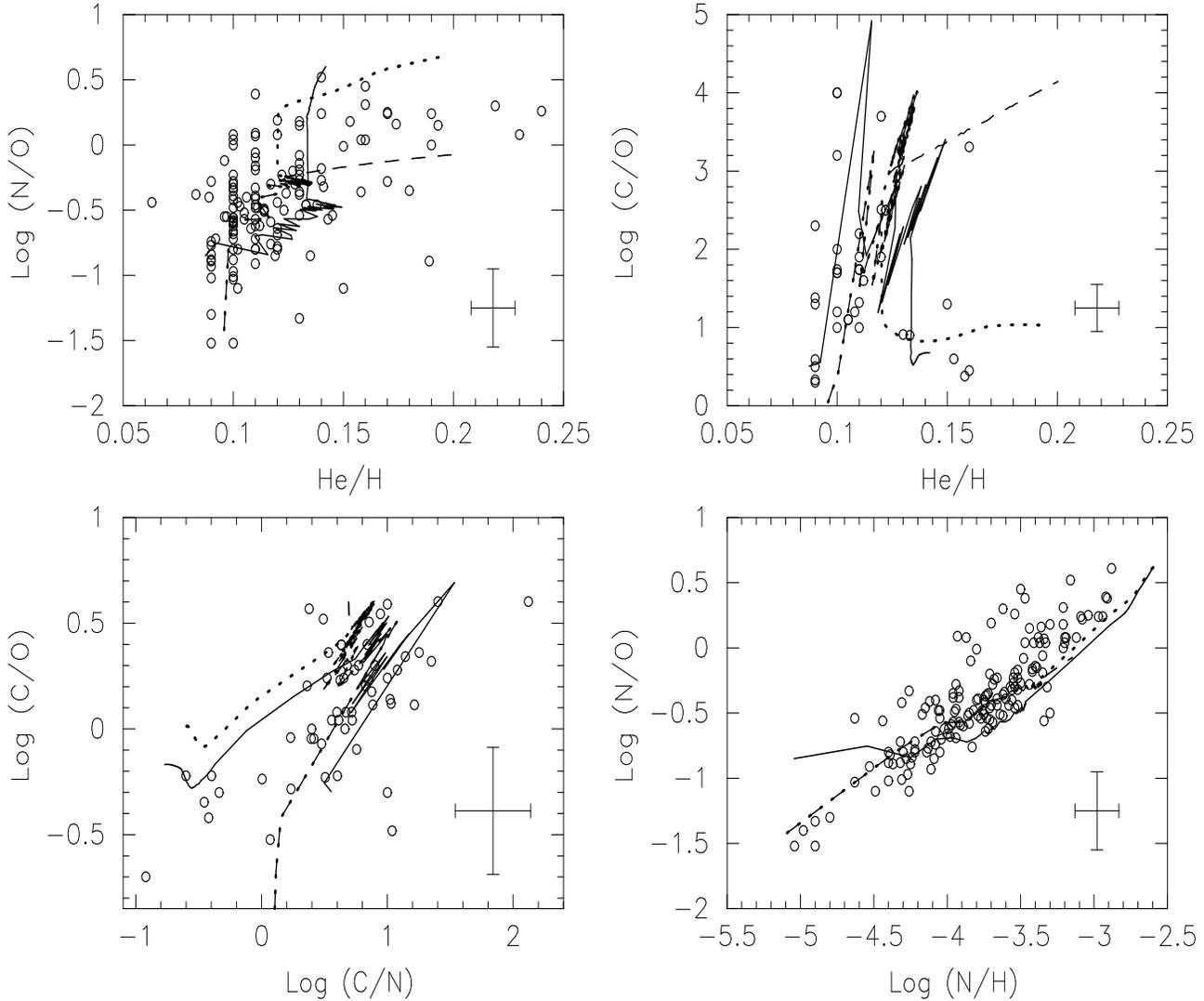,height=16.cm,width=18.cm,angle=270.}}
\caption[]{Planetary nebulae abundances (by number) predicted by the standard 
model with pre-AGB evolution according to the Geneva tracks ({\em solid 
curves}) and according to the recipes outlined in Sect. 3 ({\em dotted 
curves}). The latter model without HBB is shown for comparison ({\em dashed 
curves}). Abundances observed in PNe in the Galactic disk  are shown by 
{\em open circles} (data mainly from Aller \& Cryzack (1983), Zuckerman \& 
Aller (1986), Aller \& Keyes (1987), and Kaler \etal (1990)). Typical errors 
in the observations are indicated at the bottom right corner of each panel}
\end{figure*}

\subsection{Comparison with PN abundances observed in the Galactic disk}

In GHJ we compared the mean abundances in the envelopes of AGB stars
predicted by the standard model with the abundances observed in PNe in
the Galactic disk. Here we repeat part of this analysis with improved
model input and put emphasis on the differences in the description of
pre-AGB evolution between the Geneva models and that outlined in
GJ. In particular, we consider in more detail the effects of second
dredge-up and HBB on the predicted abundances and address the
uncertainties involved.

In the model, the abundances within PNe are estimated by averaging the
abundances in the ejecta of AGB stars over the final $\tau_{\rm PN} =$
25000 yr (\eg Pottasch 1995). We neglect any changes in the ejected
shell abundances during the post-AGB phase, \eg due to a late thermal
pulse (Sch\"{o}nberner 1983), which is expected to be a rare event, or
due to selective element depletion by dust formation. The latter
process may affect the composition both in the wind of an AGB star and
during the post-AGB phase (\eg Bond 1992; van Winckel \etal 1992) but
is neglected here for simplicity.

We assume an upper mass limit of 8 \ms for stars that ultimately can
become a PN (with final core mass less than $\sim$1.2 \mss) and ignore
the possibility that not all our model AGB stars will become PNe. In
fact, some of the low-mass AGB stars may evolve so slowly during the
post-AGB phase that the material previously collected in the wind is
dispersed before the central star has become hot enough to ionize this
material. Also, the upper mass limit for AGB stars is matter of debate
and may range between 6 and $\sim 9$ \mss, depending on the critical
mass for carbon ignition in an electron degenerate core and on details
of the stellar mass-loss scenario (\cf GJ; Vassiliadis \& Wood 1993;
Hashimoto \etal 1993). Furthermore, we assume a constant value of
$\tau_{\rm PN} = 25000$ yr. In reality, the time during which the mass
accumulated in a PN has been swept up on the AGB may depend on the
mass and initial composition of the progenitor. Nevertheless, we do
not expect that these simplifications will alter our qualitative
conclusions given below.

The observed PNe abundances are taken from various sources, \ie mainly
from Aller \& Cryzack (1983), Zuckerman \& Aller (1986), Aller \&
Keyes (1987), and Kaler \etal (1990). The few halo PNe are excluded as
the present comparison concentrates on AGB stars in the Galactic disk.
Errors in the observed abundances are typically 0.015 in He/H and
about 0.2$-$0.25 dex in all other ratios considered in Fig. 4.

The PNe nowadays observed in the Galactic disk probably originate from
AGB stars covering a wide range in initial mass, \ie $\sim$0.85$-$8
\mss.  This means that the progenitors of these PNe were formed at
galactic ages ranging from about 10$-$15 Gyr to 50 Myr ago (see \eg
Schaller \etal 1992).  Therefore, the initial element abundances of
these PN progenitors are expected to differ considerably since the
enrichment of the Galactic disk ISM over this time interval has been
substantial (\eg Twarog 1980; Edvardsson \etal 1993).  When comparing
the abundances predicted in the envelopes of final stage AGB stars
with those observed in PNe, we take this important effect into account
by incorporating a self-consistent model for the chemical evolution of
the Galactic disk (van den Hoek \etal 1996; GHJ).  Since the
metallicity dependent AGB yields and the resulting chemical evolution
of the Galactic disk are mutually dependent, an iterative solution
method was applied. The adopted star formation history (SFR) and
initial mass function (IMF) in this model were derived using
observational constraints to the abundance-abundance variations with
age of stars in the solar neighbourhood, the metallicity and age
distributions of long-living stars as well as constraints to the
current space density and formation rate of several post-main-sequence
star populations.

Resulting abundance-ratios (by number) in PNe are shown in Fig. 4 in
case of the standard model assuming pre-AGB evolution according to the
Geneva tracks (Tables 2$-$21).  We verified that the resulting
abundances are insensitive to the adopted PN lifetime up to $\tau_{\rm
PN} = $ 50000 yr.  In general, good agreement is found between the
observed and predicted PN abundances despite the uncertainties
involved.  We find that the overall trend of the observations is
reproduced well by the standard model independent of the adopted
chemical enrichment history of the Galactic disk. However, some
discrepancies are present between the standard model (with pre-AGB
evolution according to the Geneva tracks) and the observations, in
particular at large values of He/H $\ga$0.15.

For comparison, we show in Fig. 4 the PN abundances predicted by the
standard model with pre-AGB evolution according to the recipes
outlined in Sect. 3.  In this case, the enhanced effect of second
dredge-up can account for massive AGB stars with He/H up to $\sim$0.18
in their envelopes. This suggests that second dredge-up has been
relatively important at least for some of the PNe in our sample with
He/H $\la$0.15.  Alternatively, a substantial fraction of the hydrogen
contained in the outer envelope may have turned into helium.  Since
PNe may evolve from a H and/or He-shell burning AGB star, this will
determine the distribution of He/H abundance ratios observed for a
given progenitor mass. We have included in Table 32 the yields of H
and He for the standard model with second dredge-up as described by RV
(\cf Sect. 3.2) which provides reasonable agreement with the observed
PN abundances of He/H up to $\sim$0.2, in particular for the more
massive PNe.

The effect of HBB on the predicted abundances can be seen in Fig. 4 by
comparison of the standard model with $m_{\rm HBB}$ = 0.8 and 1.3 \mss
(\ie no HBB), respectively. Our results indicate that the standard
model overestimates the effect of HBB on the resulting N/O abundance
ratios in PNe with progenitors mass $\ga$5$-$6 \mss. We note that the
standard model takes into account the maximum effect of HBB as
described by RV so that values of the mixing length parameter $\alpha$
$<$2 in case of RV are probably more appropriate for massive AGB
stars.  On the other hand, models without HBB are inconsistent with
the observed N/O abundances as well as with independent observations
discussed in Sect. 4.2. Therefore, the range of N/O abundances
observed in the envelopes of post-AGB stars allows for variations in
the importance of HBB roughly covering the range from $m_{\rm HBB}$ =
0.8 to 0.9 \mss. In case of reduced HBB (\ie $m_{\rm HBB}$ = 0.9
\mss), the CNO yields of massive stars in Tables 22$-$31 are more
suitable than those given for the standard model.

The procedure to approximate the effect of HBB in a semi-analytical
way has been described in the Appendix of GJ. Here the basic
parameters were determined by fitting the RV ($\alpha$ = 2, $\eta_{\rm
AGB}$ = 0.33) case for which the effect of HBB is largest. Thus, as
the standard model has $\eta_{\rm AGB}$ = 4, possible effects of HBB
varying with in particular mass loss were neglected. In fact, the
temperature structure of the envelope is expected to change when the
number of thermal pulses decreases with increasing values of
$\eta_{\rm AGB}$. This may reduce the amount of HBB occuring in the
convective envelope and affect the resulting abundances as observed
for PNe with log (N/O) $\la$ $-0.5$ and He/H $\ga$0.15 (\cf Fig. 4).

We emphasize that the resulting abundances of PNe do depend strongly
on the initial element abundances of their progenitors, \ie are very
sensitive to the detailed chemical enrichment of the Galactic disk.  A
considerable part of the scatter observed in Fig. 4 is expected to be
caused (in addition to experimental errors) by substantial variations
in the initial abundances of PN progenitors due to the inhomogeneous
chemical evolution of the Galactic disk ISM (\eg van den Hoek \& de
Jong 1996). Furthermore, the progenitors of the PNe nowadays observed
in the solar neighbourhood probably have been formed over a large
range in galactocentric distance (\eg Wielen \etal 1996) and thus with a
large range in initial metallicity according to the radial abundance
gradients in the disk ISM (\eg Shaver \etal 1983).  Therefore, we
expect the agreement between the predicted and observed PN
abundance-ratios to improve further when averaging over a range in
initial composition for a given progenitor mass.

We conclude that the abundance-ratios predicted by the standard model
are consistent with the observed abundances in virtually all the PNe
in our sample when we allow for plausible variations in strength of
second dredge-up and HBB. 

\subsection{Final remarks}

The primary application of the stellar yields presented in this paper
is probably in galactic chemical evolution studies. 

As models with the default parameters for the mass loss on the AGB,
third dredge-up efficiency, and HBB fit many constrains in our
galaxy and the LMC (GJ/GHJ), the corresponding yields (Tables 2-21) are
probably the most appropriate ones to use. Possible alternatives are
models with less HBB (Tables 22-31), or using the pre-AGB evolution from the 
synthetic model (Table 32).
We argued in Sect 5.2 that the scaling factor $\eta_{\rm AGB}$ of the Reimers 
law may be different for low metallicities. To simulate this effect one may 
want to use the yields for models with ${\eta}_{\rm AGB}$ = 2 for $Z$ =
0.004 and ${\eta}_{\rm AGB}$ = 1 for $Z$ = 0.001 (Tables 33-38). \\

\noindent {\em Acknowledgements.} We like to thank the referee Georges Meynet for careful reading of the paper and encouraging remarks. 
It is a pleasure to thank Achim Weiss for his comments on an earlier version of this paper. 
The research of LBH and MG has been 
supported under grants 782-372-028 and 782-373-030 by the Netherlands 
Foundation for Research in Astronomy (ASTRON) which is financially supported
by the Netherlands Organisation for Scientific Research (NWO). \\

\

{}


\renewcommand{\baselinestretch}{1.}



\begin{table*}[htp]
\caption[]{
Pre-AGB   yields for Z$_{\rm ini}$= 0.001, $\eta_{\rm AGB}$=4, and $m_{\rm HBB}$=0.8  }
 \begin{flushleft}
 \begin{tabular}{lllllllllll}
 \hline
         $m_{\rm ini}$ & H & $^{4}$He & $^{12}$C & $^{13}$C 
       & $^{14}$N & 
  $^{16}$O & Y$_{\rm CNO}$ & Y$_{\rm TOT}$ & 
          $\Delta m_{\rm ej}$ & $m_{\rm end}$ \\ \hline

  .8 &  .000E+00 &  .000E+00 &  .000E+00 &  .000E+00 &  .000E+00 &  .000E+00 &  .000E+00 &  .000E+00 & 
  .00 &   .80 \\ 
  .9 & -.676E-03 &  .682E-03 & -.762E-06 & -.124E-05 &  .838E-06 &  .622E-07 & -.106E-05 & -.106E-05 & 
  .03 &   .87 \\ 
 1.0 & -.726E-03 &  .669E-03 & -.110E-05 & -.107E-05 &  .126E-05 &  .450E-07 & -.829E-06 & -.829E-06 & 
  .03 &   .97 \\ 
 1.3 & -.173E-03 &  .173E-03 & -.438E-06 & -.241E-06 &  .506E-06 &  .640E-08 & -.153E-06 & -.153E-06 & 
  .01 &  1.24 \\ 
 1.3 & -.435E-03 &  .390E-03 & -.101E-05 & -.614E-06 &  .116E-05 & -.160E-07 & -.459E-06 & -.459E-06 & 
  .02 &  1.23 \\ 
 1.5 & -.176E-03 &  .171E-03 & -.592E-06 & -.265E-06 &  .695E-06 &  .533E-08 & -.140E-06 & -.140E-06 & 
  .01 &  1.49 \\ 
 1.7 & -.388E-04 &  .391E-04 & -.166E-06 & -.658E-07 &  .197E-06 & -.294E-08 & -.322E-07 & -.322E-07 & 
  .00 &  1.70 \\ 
 2.0 & -.602E-04 &  .581E-04 & -.253E-06 & -.977E-07 &  .353E-06 & -.635E-07 & -.516E-07 & -.516E-07 & 
  .01 &  1.99 \\ 
 2.5 & -.317E-04 &  .310E-04 & -.294E-06 & -.105E-06 &  .398E-06 & -.624E-07 & -.530E-07 & -.530E-07 & 
  .01 &  2.49 \\ 
 3.0 & -.151E-04 &  .123E-04 & -.335E-06 & -.125E-06 &  .394E-06 & -.700E-08 & -.641E-07 & -.641E-07 & 
  .01 &  2.99 \\ 
 4.0 & -.420E-05 &  .490E-05 & -.231E-06 & -.100E-06 &  .274E-06 & -.600E-08 & -.580E-07 & -.580E-07 & 
  .01 &  3.99 \\ 
 5.0 & -.119E-04 &  .274E-05 & -.190E-06 & -.880E-07 &  .225E-06 & -.660E-08 & -.571E-07 & -.571E-07 & 
  .01 &  4.99 \\ 
 7.0 & -.777E-05 &  .311E-05 & -.168E-06 & -.784E-07 &  .206E-06 & -.117E-07 & -.476E-07 & -.476E-07 & 
  .02 &  6.98 \\ 
 \hline
 \end{tabular}
 \end{flushleft}
 \end{table*}

 \begin{table*}[ht]
\caption[]{
AGB       yields for Z$_{\rm ini}$= 0.001, $\eta_{\rm AGB}$=4, and $m_{\rm HBB}$=0.8  }
 \begin{flushleft}
 \begin{tabular}{lllllllllll}
 \hline
         $m_{\rm ini}$ & H & $^{4}$He & $^{12}$C & $^{13}$C 
       & $^{14}$N & 
  $^{16}$O & Y$_{\rm CNO}$ & Y$_{\rm TOT}$ & 
          $\Delta m_{\rm ej}$ & $m_{\rm end}$ \\ \hline

  .8 & -.121E-01 &  .121E-01 & -.187E-04 & -.107E-04 &  .218E-04 & -.176E-05 & -.929E-05 & -.149E-07 & 
  .27 &   .55 \\ 
  .9 & -.127E-01 &  .127E-01 & -.207E-04 & -.117E-04 &  .242E-04 & -.193E-05 & -.101E-04 & -.147E-07 & 
  .33 &   .57 \\ 
 1.0 & -.133E-01 &  .133E-01 & -.236E-04 & -.131E-04 &  .276E-04 & -.215E-05 & -.113E-04 & -.170E-07 & 
  .41 &   .59 \\ 
 1.3 & -.185E-01 &  .168E-01 &  .149E-02 & -.156E-04 &  .332E-04 &  .132E-03 &  .164E-02 &  .166E-02 & 
  .60 &   .65 \\ 
 1.5 & -.196E-01 &  .172E-01 &  .221E-02 & -.173E-04 &  .380E-04 &  .196E-03 &  .243E-02 &  .244E-02 & 
  .79 &   .71 \\ 
 1.7 & -.180E-01 &  .154E-01 &  .230E-02 & -.174E-04 &  .396E-04 &  .204E-03 &  .252E-02 &  .254E-02 & 
  .91 &   .79 \\ 
 2.0 & -.222E-01 &  .187E-01 &  .305E-02 & -.188E-04 &  .701E-04 &  .262E-03 &  .337E-02 &  .354E-02 & 
 1.13 &   .86 \\
\hline
 \end{tabular}
 \end{flushleft}
 \end{table*}

 \begin{table*}[thp]
\caption[]{
Final AGB yields for Z$_{\rm ini}$= 0.001, $\eta_{\rm AGB}$=4, and $m_{\rm HBB}$=0.8  }
 \begin{flushleft}
 \begin{tabular}{lllllllllll}
 \hline
         $m_{\rm ini}$ & H & $^{4}$He & $^{12}$C & $^{13}$C 
       & $^{14}$N & 
  $^{16}$O & Y$_{\rm CNO}$ & Y$_{\rm TOT}$ & 
          $\Delta m_{\rm ej}$ & $m_{\rm end}$ \\ \hline

  .8 & -.374E-03 &  .374E-03 & -.579E-06 & -.330E-06 &  .676E-06 & -.546E-07 & -.288E-06 & -.436E-09 & 
  .01 &   .54 \\ 
  .9 & -.689E-03 &  .689E-03 & -.113E-05 & -.634E-06 &  .132E-05 & -.105E-06 & -.551E-06 & -.889E-09 & 
  .02 &   .56 \\ 
 1.0 & -.744E-03 &  .744E-03 & -.132E-05 & -.728E-06 &  .154E-05 & -.120E-06 & -.628E-06 & -.100E-08 & 
  .02 &   .57 \\ 
 1.3 & -.576E-02 &  .433E-02 &  .131E-02 & -.166E-05 &  .271E-05 &  .116E-03 &  .143E-02 &  .143E-02 & 
  .06 &   .59 \\ 
 1.5 & -.468E-02 &  .360E-02 &  .997E-03 & -.194E-05 &  .371E-05 &  .883E-04 &  .109E-02 &  .109E-02 & 
  .09 &   .62 \\ 
 1.7 & -.549E-02 &  .426E-02 &  .113E-02 & -.294E-05 &  .610E-05 &  .999E-04 &  .123E-02 &  .123E-02 & 
  .15 &   .64 \\ 
 2.0 & -.725E-02 &  .550E-02 &  .158E-02 & -.313E-05 &  .106E-04 &  .138E-03 &  .172E-02 &  .175E-02 & 
  .19 &   .68 \\
\hline
 \end{tabular}
 \end{flushleft}
 \end{table*}

 \begin{table*}[htp]
\caption[]{
Total     yields for Z$_{\rm ini}$= 0.001, $\eta_{\rm AGB}$=4, and $m_{\rm HBB}$=0.8  }
 \begin{flushleft}
 \begin{tabular}{lllllllllll}
 \hline
         $m_{\rm ini}$ & H & $^{4}$He & $^{12}$C & $^{13}$C 
       & $^{14}$N & 
  $^{16}$O & Y$_{\rm CNO}$ & Y$_{\rm TOT}$ & 
          $\Delta m_{\rm ej}$ & $m_{\rm end}$ \\ \hline

  .9 & -.140E-01 &  .140E-01 & -.226E-04 & -.135E-04 &  .264E-04 & -.197E-05 & -.117E-04 & -.108E-05 & 
  .38 &   .56 \\ 
 1.0 & -.148E-01 &  .148E-01 & -.260E-04 & -.149E-04 &  .304E-04 & -.223E-05 & -.127E-04 & -.847E-06 & 
  .46 &   .57 \\ 
 1.3 & -.244E-01 &  .213E-01 &  .281E-02 & -.175E-04 &  .364E-04 &  .249E-03 &  .308E-02 &  .309E-02 & 
  .67 &   .59 \\ 
 1.3 & -.247E-01 &  .215E-01 &  .281E-02 & -.178E-04 &  .371E-04 &  .249E-03 &  .308E-02 &  .309E-02 & 
  .69 &   .59 \\ 
 1.5 & -.245E-01 &  .210E-01 &  .321E-02 & -.195E-04 &  .424E-04 &  .284E-03 &  .351E-02 &  .353E-02 & 
  .89 &   .62 \\ 
 1.7 & -.235E-01 &  .197E-01 &  .343E-02 & -.204E-04 &  .459E-04 &  .304E-03 &  .375E-02 &  .377E-02 & 
 1.06 &   .64 \\ 
 2.0 & -.295E-01 &  .242E-01 &  .463E-02 & -.220E-04 &  .811E-04 &  .400E-03 &  .509E-02 &  .529E-02 & 
 1.32 &   .68 \\ 
 2.5 & -.244E-01 &  .188E-01 &  .495E-02 & -.227E-04 &  .882E-04 &  .429E-03 &  .544E-02 &  .559E-02 & 
 1.76 &   .74 \\ 
 3.0 & -.203E-01 &  .149E-01 &  .472E-02 & -.234E-04 &  .802E-04 &  .417E-03 &  .519E-02 &  .535E-02 & 
 2.18 &   .82 \\ 
 4.0 & -.268E-01 &  .202E-01 &  .109E-02 &  .223E-03 &  .472E-02 &  .431E-03 &  .647E-02 &  .656E-02 & 
 3.09 &   .91 \\ 
 5.0 & -.311E-01 &  .234E-01 &  .580E-03 &  .525E-04 &  .657E-02 &  .412E-03 &  .761E-02 &  .772E-02 & 
 4.02 &   .98 \\ 
 7.0 & -.412E-01 &  .316E-01 &  .734E-03 &  .910E-04 &  .869E-02 &  .629E-04 &  .958E-02 &  .961E-02 & 
 5.87 &  1.13 \\ 
 8.0 & -.382E-01 &  .295E-01 &  .603E-03 &  .858E-04 &  .794E-02 &  .960E-04 &  .872E-02 &  .878E-02 & 
 6.17 &  1.84 \\ 
 \hline
 \end{tabular}
 \end{flushleft}
 \end{table*}

 \begin{table*}[htp]
\caption[]{
Pre-AGB   yields for Z$_{\rm ini}$= 0.004, $\eta_{\rm AGB}$=4, and $m_{\rm HBB}$=0.8  }
 \begin{flushleft}
 \begin{tabular}{lllllllllll}
 \hline
         $m_{\rm ini}$ & H & $^{4}$He & $^{12}$C & $^{13}$C 
       & $^{14}$N & 
  $^{16}$O & Y$_{\rm CNO}$ & Y$_{\rm TOT}$ & 
          $\Delta m_{\rm ej}$ & $m_{\rm end}$ \\ \hline

  .8 & -.480E-04 &  .385E-04 &  .128E-05 & -.205E-07 &  .330E-06 &  .276E-05 &  .436E-05 &  .436E-05 & 
  .00 &   .80 \\ 
  .9 & -.132E-03 &  .114E-03 &  .291E-05 & -.189E-07 &  .109E-05 &  .684E-05 &  .108E-04 &  .108E-04 & 
  .00 &   .90 \\ 
 1.0 & -.178E-03 &  .154E-03 &  .337E-05 & -.220E-07 &  .176E-05 &  .868E-05 &  .138E-04 &  .138E-04 & 
  .01 &   .99 \\ 
 1.3 & -.883E-04 &  .774E-04 &  .140E-05 & -.136E-07 &  .118E-05 &  .429E-05 &  .687E-05 &  .687E-05 & 
  .00 &  1.25 \\ 
 1.3 & -.704E-04 &  .648E-04 &  .127E-05 & -.720E-08 &  .100E-05 &  .378E-05 &  .606E-05 &  .606E-05 & 
  .00 &  1.25 \\ 
 1.5 & -.575E-03 &  .507E-03 &  .788E-05 & -.109E-06 &  .960E-05 &  .286E-04 &  .461E-04 &  .461E-04 & 
  .03 &  1.47 \\ 
 1.7 & -.455E-03 &  .405E-03 &  .586E-05 & -.102E-06 &  .863E-05 &  .224E-04 &  .369E-04 &  .369E-04 & 
  .02 &  1.68 \\ 
 2.0 & -.243E-03 &  .215E-03 &  .291E-05 & -.590E-07 &  .470E-05 &  .106E-04 &  .182E-04 &  .182E-04 & 
  .01 &  1.99 \\ 
 2.5 & -.255E-03 &  .238E-03 &  .275E-05 & -.512E-07 &  .564E-05 &  .106E-04 &  .189E-04 &  .189E-04 & 
  .02 &  2.48 \\ 
 3.0 & -.340E-03 &  .295E-03 &  .400E-05 & -.660E-07 &  .827E-05 &  .154E-04 &  .276E-04 &  .276E-04 & 
  .03 &  2.97 \\ 
 4.0 & -.141E-03 &  .129E-03 &  .315E-05 & -.355E-07 &  .523E-05 &  .115E-04 &  .199E-04 &  .199E-04 & 
  .03 &  3.97 \\ 
 5.0 & -.926E-04 &  .761E-04 &  .226E-05 & -.196E-07 &  .344E-05 &  .806E-05 &  .138E-04 &  .138E-04 & 
  .03 &  4.97 \\ 
 7.0 & -.192E-03 &  .160E-03 &  .490E-05 & -.431E-07 &  .740E-05 &  .172E-04 &  .296E-04 &  .296E-04 & 
  .08 &  6.92 \\ 
 \hline
 \end{tabular}
 \end{flushleft}
 \end{table*}

 \begin{table*}[htp]
\caption[]{
AGB       yields for Z$_{\rm ini}$= 0.004, $\eta_{\rm AGB}$=4, and $m_{\rm HBB}$=0.8  }
 \begin{flushleft}
 \begin{tabular}{lllllllllll}
 \hline
         $m_{\rm ini}$ & H & $^{4}$He & $^{12}$C & $^{13}$C 
       & $^{14}$N & 
  $^{16}$O & Y$_{\rm CNO}$ & Y$_{\rm TOT}$ & 
          $\Delta m_{\rm ej}$ & $m_{\rm end}$ \\ \hline

  .8 & -.108E-01 &  .108E-01 & -.753E-04 &  .564E-05 &  .879E-04 & -.649E-05 &  .117E-04 & -.144E-07 & 
  .25 &   .57 \\ 
  .9 & -.110E-01 &  .110E-01 & -.811E-04 &  .593E-05 &  .947E-04 & -.689E-05 &  .126E-04 &  .210E-09 & 
  .29 &   .61 \\ 
 1.0 & -.119E-01 &  .119E-01 & -.950E-04 &  .673E-05 &  .111E-03 & -.791E-05 &  .147E-04 & -.698E-08 & 
  .38 &   .62 \\ 
 1.3 & -.155E-01 &  .144E-01 &  .890E-03 &  .770E-05 &  .138E-03 &  .730E-04 &  .111E-02 &  .109E-02 & 
  .57 &   .68 \\ 
 1.5 & -.176E-01 &  .157E-01 &  .164E-02 &  .864E-05 &  .167E-03 &  .135E-03 &  .195E-02 &  .193E-02 & 
  .80 &   .70 \\ 
 1.7 & -.205E-01 &  .174E-01 &  .272E-02 &  .941E-05 &  .194E-03 &  .224E-03 &  .315E-02 &  .311E-02 & 
 1.03 &   .67 \\ 
 2.0 & -.288E-01 &  .239E-01 &  .411E-02 &  .103E-04 &  .314E-03 &  .287E-03 &  .472E-02 &  .493E-02 & 
 1.29 &   .70 \\
\hline
 \end{tabular}
 \end{flushleft}
 \end{table*}

 \begin{table*}[htp]
\caption[]{
Final AGB yields for Z$_{\rm ini}$= 0.004, $\eta_{\rm AGB}$=4, and $m_{\rm HBB}$=0.8  }
 \begin{flushleft}
 \begin{tabular}{lllllllllll}
 \hline
         $m_{\rm ini}$ & H & $^{4}$He & $^{12}$C & $^{13}$C 
       & $^{14}$N & 
  $^{16}$O & Y$_{\rm CNO}$ & Y$_{\rm TOT}$ & 
          $\Delta m_{\rm ej}$ & $m_{\rm end}$ \\ \hline

  .8 & -.545E-04 &  .545E-04 & -.380E-06 &  .284E-07 &  .443E-06 & -.327E-07 &  .591E-07 & -.725E-10 & 
  .00 &   .57 \\ 
  .9 & -.107E-02 &  .107E-02 & -.788E-05 &  .577E-06 &  .920E-05 & -.670E-06 &  .122E-05 &  .437E-09 & 
  .03 &   .58 \\ 
 1.0 & -.397E-02 &  .313E-02 &  .769E-03 &  .688E-06 &  .114E-04 &  .628E-04 &  .844E-03 &  .840E-03 & 
  .05 &   .58 \\ 
 1.3 & -.478E-02 &  .388E-02 &  .815E-03 &  .119E-05 &  .214E-04 &  .666E-04 &  .905E-03 &  .899E-03 & 
  .10 &   .59 \\ 
 1.5 & -.334E-02 &  .274E-02 &  .534E-03 &  .107E-05 &  .208E-04 &  .438E-04 &  .600E-03 &  .596E-03 & 
  .10 &   .60 \\ 
 1.7 & -.335E-02 &  .251E-02 &  .771E-03 &  .564E-06 &  .119E-04 &  .631E-04 &  .847E-03 &  .843E-03 & 
  .07 &   .60 \\ 
 2.0 & -.368E-02 &  .277E-02 &  .813E-03 &  .593E-06 &  .187E-04 &  .634E-04 &  .896E-03 &  .907E-03 & 
  .08 &   .62 \\
\hline
 \end{tabular}
 \end{flushleft}
 \end{table*}

 \begin{table*}[htp]
\caption[]{
Total     yields for Z$_{\rm ini}$= 0.004, $\eta_{\rm AGB}$=4, and $m_{\rm HBB}$=0.8  }
 \begin{flushleft}
 \begin{tabular}{lllllllllll}
 \hline
         $m_{\rm ini}$ & H & $^{4}$He & $^{12}$C & $^{13}$C 
       & $^{14}$N & 
  $^{16}$O & Y$_{\rm CNO}$ & Y$_{\rm TOT}$ & 
          $\Delta m_{\rm ej}$ & $m_{\rm end}$ \\ \hline

  .9 & -.122E-01 &  .122E-01 & -.861E-04 &  .649E-05 &  .105E-03 & -.722E-06 &  .247E-04 &  .108E-04 & 
  .33 &   .58 \\ 
 1.0 & -.161E-01 &  .152E-01 &  .678E-03 &  .739E-05 &  .124E-03 &  .636E-04 &  .873E-03 &  .854E-03 & 
  .43 &   .58 \\ 
 1.3 & -.203E-01 &  .184E-01 &  .171E-02 &  .888E-05 &  .160E-03 &  .144E-03 &  .202E-02 &  .199E-02 & 
  .66 &   .59 \\ 
 1.3 & -.203E-01 &  .183E-01 &  .171E-02 &  .888E-05 &  .160E-03 &  .143E-03 &  .202E-02 &  .199E-02 & 
  .66 &   .59 \\ 
 1.5 & -.215E-01 &  .189E-01 &  .219E-02 &  .961E-05 &  .198E-03 &  .207E-03 &  .260E-02 &  .257E-02 & 
  .93 &   .60 \\ 
 1.7 & -.243E-01 &  .203E-01 &  .350E-02 &  .988E-05 &  .215E-03 &  .309E-03 &  .403E-02 &  .399E-02 & 
 1.12 &   .60 \\ 
 2.0 & -.328E-01 &  .269E-01 &  .493E-02 &  .108E-04 &  .337E-03 &  .361E-03 &  .564E-02 &  .586E-02 & 
 1.38 &   .62 \\ 
 2.5 & -.378E-01 &  .309E-01 &  .590E-02 &  .123E-04 &  .418E-03 &  .396E-03 &  .673E-02 &  .682E-02 & 
 1.86 &   .64 \\ 
 3.0 & -.402E-01 &  .318E-01 &  .738E-02 &  .126E-04 &  .444E-03 &  .512E-03 &  .835E-02 &  .843E-02 & 
 2.34 &   .66 \\ 
 4.0 & -.257E-01 &  .199E-01 &  .841E-03 &  .267E-03 &  .447E-02 &  .276E-03 &  .585E-02 &  .587E-02 & 
 3.09 &   .91 \\ 
 5.0 & -.317E-01 &  .247E-01 &  .294E-03 &  .878E-04 &  .639E-02 &  .257E-03 &  .703E-02 &  .707E-02 & 
 4.02 &   .98 \\ 
 7.0 & -.371E-01 &  .282E-01 &  .386E-03 &  .123E-03 &  .858E-02 & -.254E-03 &  .884E-02 &  .888E-02 & 
 5.88 &  1.12 \\ 
 8.0 & -.422E-01 &  .326E-01 &  .328E-03 &  .133E-03 &  .939E-02 & -.321E-03 &  .953E-02 &  .957E-02 & 
 6.81 &  1.20 \\ 
 \hline
 \end{tabular}
 \end{flushleft}
 \end{table*}

 \begin{table*}[htp]
\caption[]{
Pre-AGB   yields for Z$_{\rm ini}$= 0.008, $\eta_{\rm AGB}$=4, and $m_{\rm HBB}$=0.8  }
 \begin{flushleft}
 \begin{tabular}{lllllllllll}
 \hline
         $m_{\rm ini}$ & H & $^{4}$He & $^{12}$C & $^{13}$C 
       & $^{14}$N & 
  $^{16}$O & Y$_{\rm CNO}$ & Y$_{\rm TOT}$ & 
          $\Delta m_{\rm ej}$ & $m_{\rm end}$ \\ \hline

  .8 & -.591E-03 &  .491E-03 &  .196E-04 &  .250E-06 &  .683E-05 &  .482E-04 &  .751E-04 &  .751E-04 & 
  .01 &   .79 \\ 
  .9 & -.571E-03 &  .471E-03 &  .166E-04 &  .293E-06 &  .748E-05 &  .438E-04 &  .683E-04 &  .683E-04 & 
  .01 &   .89 \\ 
 1.0 & -.102E-02 &  .814E-03 &  .252E-04 &  .482E-06 &  .145E-04 &  .713E-04 &  .112E-03 &  .112E-03 & 
  .02 &   .98 \\ 
 1.3 & -.664E-03 &  .582E-03 &  .160E-04 &  .336E-06 &  .133E-04 &  .519E-04 &  .817E-04 &  .817E-04 & 
  .02 &  1.23 \\ 
 1.3 & -.777E-03 &  .657E-03 &  .186E-04 &  .418E-06 &  .151E-04 &  .600E-04 &  .943E-04 &  .943E-04 & 
  .02 &  1.23 \\ 
 1.5 & -.477E-03 &  .387E-03 &  .103E-04 &  .243E-06 &  .110E-04 &  .373E-04 &  .589E-04 &  .589E-04 & 
  .02 &  1.48 \\ 
 1.7 & -.540E-03 &  .434E-03 &  .112E-04 &  .268E-06 &  .138E-04 &  .421E-04 &  .676E-04 &  .676E-04 & 
  .02 &  1.68 \\ 
 2.0 & -.387E-03 &  .330E-03 &  .804E-05 &  .160E-06 &  .106E-04 &  .292E-04 &  .482E-04 &  .482E-04 & 
  .02 &  1.98 \\ 
 2.5 & -.495E-03 &  .420E-03 &  .827E-05 &  .199E-06 &  .136E-04 &  .310E-04 &  .532E-04 &  .532E-04 & 
  .02 &  2.48 \\ 
 3.0 & -.511E-03 &  .448E-03 &  .922E-05 &  .232E-06 &  .155E-04 &  .343E-04 &  .594E-04 &  .594E-04 & 
  .03 &  2.97 \\ 
 4.0 & -.457E-03 &  .392E-03 &  .978E-05 &  .261E-06 &  .154E-04 &  .357E-04 &  .613E-04 &  .613E-04 & 
  .04 &  3.96 \\ 
 5.0 & -.444E-03 &  .354E-03 &  .100E-04 &  .274E-06 &  .155E-04 &  .362E-04 &  .620E-04 &  .620E-04 & 
  .05 &  4.95 \\ 
 7.0 & -.699E-03 &  .573E-03 &  .156E-04 &  .437E-06 &  .246E-04 &  .558E-04 &  .966E-04 &  .966E-04 & 
  .12 &  6.88 \\ 
 \hline
 \end{tabular}
 \end{flushleft}
 \end{table*}

 \begin{table*}[htp]
\caption[]{
AGB       yields for Z$_{\rm ini}$= 0.008, $\eta_{\rm AGB}$=4, and $m_{\rm HBB}$=0.8  }
 \begin{flushleft}
 \begin{tabular}{lllllllllll}
 \hline
         $m_{\rm ini}$ & H & $^{4}$He & $^{12}$C & $^{13}$C 
       & $^{14}$N & 
  $^{16}$O & Y$_{\rm CNO}$ & Y$_{\rm TOT}$ & 
          $\Delta m_{\rm ej}$ & $m_{\rm end}$ \\ \hline

  .8 & -.110E-01 &  .110E-01 & -.147E-03 &  .111E-04 &  .172E-03 & -.139E-04 &  .218E-04 &  .225E-08 & 
  .27 &   .56 \\ 
  .9 & -.101E-01 &  .101E-01 & -.144E-03 &  .106E-04 &  .168E-03 & -.134E-04 &  .212E-04 &  .175E-07 & 
  .28 &   .62 \\ 
 1.0 & -.111E-01 &  .111E-01 & -.171E-03 &  .122E-04 &  .200E-03 & -.156E-04 &  .252E-04 &  .666E-08 & 
  .37 &   .63 \\ 
 1.3 & -.141E-01 &  .132E-01 &  .610E-03 &  .142E-04 &  .251E-03 &  .408E-04 &  .916E-03 &  .879E-03 & 
  .56 &   .69 \\ 
 1.5 & -.157E-01 &  .141E-01 &  .121E-02 &  .160E-04 &  .305E-03 &  .842E-04 &  .161E-02 &  .156E-02 & 
  .79 &   .71 \\ 
 1.7 & -.185E-01 &  .158E-01 &  .220E-02 &  .176E-04 &  .358E-03 &  .156E-03 &  .273E-02 &  .267E-02 & 
 1.03 &   .67 \\ 
 2.0 & -.257E-01 &  .214E-01 &  .346E-02 &  .198E-04 &  .525E-03 &  .165E-03 &  .417E-02 &  .435E-02 & 
 1.27 &   .72 \\
\hline
 \end{tabular}
 \end{flushleft}
 \end{table*}

 \begin{table*}[htp]
\caption[]{
Final AGB yields for Z$_{\rm ini}$= 0.008, $\eta_{\rm AGB}$=4, and $m_{\rm HBB}$=0.8  }
 \begin{flushleft}
 \begin{tabular}{lllllllllll}
 \hline
         $m_{\rm ini}$ & H & $^{4}$He & $^{12}$C & $^{13}$C 
       & $^{14}$N & 
  $^{16}$O & Y$_{\rm CNO}$ & Y$_{\rm TOT}$ & 
          $\Delta m_{\rm ej}$ & $m_{\rm end}$ \\ \hline

  .8 &  .000E+00 &  .000E+00 &  .000E+00 &  .000E+00 &  .000E+00 &  .000E+00 &  .000E+00 &  .000E+00 & 
  .00 &   .56 \\ 
  .9 & -.139E-02 &  .139E-02 & -.198E-04 &  .146E-05 &  .231E-04 & -.185E-05 &  .293E-05 &  .220E-08 & 
  .04 &   .58 \\ 
 1.0 & -.149E-02 &  .149E-02 & -.230E-04 &  .165E-05 &  .268E-04 & -.210E-05 &  .338E-05 & -.844E-09 & 
  .05 &   .58 \\ 
 1.3 & -.428E-02 &  .354E-02 &  .655E-03 &  .239E-05 &  .421E-04 &  .467E-04 &  .746E-03 &  .736E-03 & 
  .10 &   .59 \\ 
 1.5 & -.604E-02 &  .460E-02 &  .132E-02 &  .210E-05 &  .400E-04 &  .946E-04 &  .145E-02 &  .144E-02 & 
  .12 &   .59 \\ 
 1.7 & -.310E-02 &  .232E-02 &  .707E-03 &  .109E-05 &  .223E-04 &  .509E-04 &  .782E-03 &  .774E-03 & 
  .07 &   .60 \\ 
 2.0 & -.422E-02 &  .319E-02 &  .923E-03 &  .151E-05 &  .407E-04 &  .598E-04 &  .103E-02 &  .104E-02 & 
  .11 &   .61 \\
\hline
 \end{tabular}
 \end{flushleft}
 \end{table*}

 \begin{table*}[htp]
\caption[]{
Total     yields for Z$_{\rm ini}$= 0.008, $\eta_{\rm AGB}$=4, and $m_{\rm HBB}$=0.8  }
 \begin{flushleft}
 \begin{tabular}{lllllllllll}
 \hline
         $m_{\rm ini}$ & H & $^{4}$He & $^{12}$C & $^{13}$C 
       & $^{14}$N & 
  $^{16}$O & Y$_{\rm CNO}$ & Y$_{\rm TOT}$ & 
          $\Delta m_{\rm ej}$ & $m_{\rm end}$ \\ \hline

  .9 & -.121E-01 &  .120E-01 & -.147E-03 &  .124E-04 &  .198E-03 &  .285E-04 &  .925E-04 &  .683E-04 & 
  .33 &   .58 \\ 
 1.0 & -.136E-01 &  .134E-01 & -.169E-03 &  .144E-04 &  .241E-03 &  .536E-04 &  .140E-03 &  .112E-03 & 
  .44 &   .58 \\ 
 1.3 & -.191E-01 &  .174E-01 &  .128E-02 &  .170E-04 &  .306E-03 &  .139E-03 &  .174E-02 &  .170E-02 & 
  .68 &   .59 \\ 
 1.3 & -.192E-01 &  .174E-01 &  .128E-02 &  .171E-04 &  .308E-03 &  .148E-03 &  .176E-02 &  .171E-02 & 
  .68 &   .59 \\ 
 1.5 & -.222E-01 &  .191E-01 &  .253E-02 &  .183E-04 &  .356E-03 &  .216E-03 &  .312E-02 &  .306E-02 & 
  .92 &   .59 \\ 
 1.7 & -.221E-01 &  .186E-01 &  .291E-02 &  .190E-04 &  .394E-03 &  .249E-03 &  .358E-02 &  .351E-02 & 
 1.12 &   .60 \\ 
 2.0 & -.303E-01 &  .249E-01 &  .439E-02 &  .215E-04 &  .576E-03 &  .254E-03 &  .524E-02 &  .543E-02 & 
 1.39 &   .61 \\ 
 2.5 & -.391E-01 &  .325E-01 &  .553E-02 &  .236E-04 &  .733E-03 &  .235E-03 &  .653E-02 &  .658E-02 & 
 1.87 &   .63 \\ 
 3.0 & -.422E-01 &  .342E-01 &  .681E-02 &  .245E-04 &  .784E-03 &  .305E-03 &  .793E-02 &  .798E-02 & 
 2.36 &   .64 \\ 
 4.0 & -.316E-01 &  .262E-01 &  .658E-03 &  .300E-03 &  .443E-02 &  .572E-04 &  .544E-02 &  .544E-02 & 
 3.10 &   .90 \\ 
 5.0 & -.331E-01 &  .265E-01 &  .104E-05 &  .971E-04 &  .643E-02 &  .660E-04 &  .659E-02 &  .659E-02 & 
 4.03 &   .97 \\ 
 7.0 & -.395E-01 &  .311E-01 &  .458E-04 &  .130E-03 &  .881E-02 & -.632E-03 &  .835E-02 &  .835E-02 & 
 5.89 &  1.11 \\ 
 8.0 & -.508E-01 &  .406E-01 & -.520E-04 &  .158E-03 &  .111E-01 & -.992E-03 &  .102E-01 &  .102E-01 & 
 6.81 &  1.20 \\ 
 \hline
 \end{tabular}
 \end{flushleft}
 \end{table*}

 \begin{table*}[htp]
\caption[]{
Pre-AGB   yields for Z$_{\rm ini}$= 0.020, $\eta_{\rm AGB}$=4, and $m_{\rm HBB}$=0.8  }
 \begin{flushleft}
 \begin{tabular}{lllllllllll}
 \hline
         $m_{\rm ini}$ & H & $^{4}$He & $^{12}$C & $^{13}$C 
       & $^{14}$N & 
  $^{16}$O & Y$_{\rm CNO}$ & Y$_{\rm TOT}$ & 
          $\Delta m_{\rm ej}$ & $m_{\rm end}$ \\ \hline

  .8 &  .000E+00 &  .000E+00 &  .000E+00 &  .000E+00 &  .000E+00 &  .000E+00 &  .000E+00 &  .000E+00 & 
  .00 &   .80 \\ 
  .9 &  .000E+00 &  .000E+00 &  .000E+00 &  .000E+00 &  .000E+00 &  .000E+00 &  .000E+00 &  .000E+00 & 
  .00 &   .90 \\ 
 1.0 & -.100E-01 &  .781E-02 &  .371E-03 &  .124E-04 &  .197E-03 &  .103E-02 &  .161E-02 &  .161E-02 & 
  .10 &   .90 \\ 
 1.3 & -.181E-01 &  .144E-01 &  .614E-03 &  .239E-04 &  .465E-03 &  .192E-02 &  .303E-02 &  .303E-02 & 
  .24 &  1.01 \\ 
 1.3 & -.440E-02 &  .349E-02 &  .153E-03 &  .595E-05 &  .111E-03 &  .473E-03 &  .744E-03 &  .744E-03 & 
  .06 &  1.19 \\ 
 1.5 & -.334E-02 &  .264E-02 &  .110E-03 &  .453E-05 &  .994E-04 &  .371E-03 &  .586E-03 &  .586E-03 & 
  .06 &  1.44 \\ 
 1.7 & -.143E-02 &  .112E-02 &  .447E-04 &  .189E-05 &  .453E-04 &  .155E-03 &  .248E-03 &  .248E-03 & 
  .03 &  1.67 \\ 
 2.0 & -.143E-02 &  .111E-02 &  .428E-04 &  .174E-05 &  .466E-04 &  .147E-03 &  .239E-03 &  .239E-03 & 
  .03 &  1.97 \\ 
 2.5 & -.119E-02 &  .962E-03 &  .342E-04 &  .145E-05 &  .415E-04 &  .118E-03 &  .196E-03 &  .196E-03 & 
  .03 &  2.47 \\ 
 3.0 & -.234E-02 &  .190E-02 &  .645E-04 &  .282E-05 &  .840E-04 &  .222E-03 &  .375E-03 &  .375E-03 & 
  .07 &  2.93 \\ 
 4.0 & -.127E-02 &  .101E-02 &  .354E-04 &  .157E-05 &  .459E-04 &  .120E-03 &  .203E-03 &  .203E-03 & 
  .05 &  3.95 \\ 
 5.0 & -.145E-02 &  .116E-02 &  .419E-04 &  .188E-05 &  .539E-04 &  .141E-03 &  .239E-03 &  .239E-03 & 
  .08 &  4.92 \\ 
 7.0 & -.274E-02 &  .221E-02 &  .799E-04 &  .366E-05 &  .104E-03 &  .265E-03 &  .453E-03 &  .453E-03 & 
  .20 &  6.80 \\ 
 \hline
 \end{tabular}
 \end{flushleft}
 \end{table*}

 \begin{table*}[htp]
\caption[]{
AGB       yields for Z$_{\rm ini}$= 0.020, $\eta_{\rm AGB}$=4, and $m_{\rm HBB}$=0.8  }
 \begin{flushleft}
 \begin{tabular}{lllllllllll}
 \hline
         $m_{\rm ini}$ & H & $^{4}$He & $^{12}$C & $^{13}$C 
       & $^{14}$N & 
  $^{16}$O & Y$_{\rm CNO}$ & Y$_{\rm TOT}$ & 
          $\Delta m_{\rm ej}$ & $m_{\rm end}$ \\ \hline

  .8 & -.986E-02 &  .986E-02 & -.389E-03 &  .296E-04 &  .454E-03 & -.368E-04 &  .579E-04 &  .563E-09 & 
  .29 &   .54 \\ 
  .9 & -.853E-02 &  .853E-02 & -.357E-03 &  .266E-04 &  .417E-03 & -.333E-04 &  .530E-04 & -.139E-07 & 
  .28 &   .62 \\ 
 1.0 & -.918E-02 &  .918E-02 & -.418E-03 &  .301E-04 &  .488E-03 & -.382E-04 &  .617E-04 & -.244E-07 & 
  .36 &   .64 \\ 
 1.3 & -.937E-02 &  .937E-02 & -.537E-03 &  .357E-04 &  .627E-03 & -.468E-04 &  .787E-04 &  .557E-07 & 
  .55 &   .70 \\ 
 1.5 & -.986E-02 &  .948E-02 & -.268E-03 &  .398E-04 &  .755E-03 & -.375E-04 &  .489E-03 &  .389E-03 & 
  .77 &   .73 \\ 
 1.7 & -.112E-01 &  .101E-01 &  .351E-03 &  .429E-04 &  .864E-03 & -.123E-04 &  .125E-02 &  .112E-02 & 
  .97 &   .73 \\ 
 2.0 & -.176E-01 &  .151E-01 &  .132E-02 &  .510E-04 &  .108E-02 & -.914E-04 &  .235E-02 &  .252E-02 & 
 1.20 &   .77 \\
\hline
 \end{tabular}
 \end{flushleft}
 \end{table*}

 \begin{table*}[htp]
\caption[]{
Final AGB yields for Z$_{\rm ini}$= 0.020, $\eta_{\rm AGB}$=4, and $m_{\rm HBB}$=0.8  }
 \begin{flushleft}
 \begin{tabular}{lllllllllll}
 \hline
         $m_{\rm ini}$ & H & $^{4}$He & $^{12}$C & $^{13}$C 
       & $^{14}$N & 
  $^{16}$O & Y$_{\rm CNO}$ & Y$_{\rm TOT}$ & 
          $\Delta m_{\rm ej}$ & $m_{\rm end}$ \\ \hline

  .8 &  .000E+00 &  .000E+00 &  .000E+00 &  .000E+00 &  .000E+00 &  .000E+00 &  .000E+00 &  .000E+00 & 
  .00 &   .54 \\ 
  .9 & -.139E-02 &  .139E-02 & -.582E-04 &  .433E-05 &  .679E-04 & -.542E-05 &  .863E-05 & -.110E-08 & 
  .05 &   .57 \\ 
 1.0 & -.165E-02 &  .165E-02 & -.753E-04 &  .542E-05 &  .878E-04 & -.687E-05 &  .111E-04 &  .396E-08 & 
  .06 &   .58 \\ 
 1.3 & -.193E-02 &  .193E-02 & -.110E-03 &  .735E-05 &  .129E-03 & -.963E-05 &  .162E-04 & -.801E-08 & 
  .11 &   .59 \\ 
 1.5 & -.500E-02 &  .390E-02 &  .962E-03 &  .696E-05 &  .132E-03 &  .372E-04 &  .114E-02 &  .110E-02 & 
  .15 &   .59 \\ 
 1.7 & -.415E-02 &  .315E-02 &  .881E-03 &  .557E-05 &  .113E-03 &  .346E-04 &  .103E-02 &  .100E-02 & 
  .14 &   .59 \\ 
 2.0 & -.503E-02 &  .386E-02 &  .991E-03 &  .698E-05 &  .148E-03 &  .220E-04 &  .117E-02 &  .118E-02 & 
  .18 &   .60 \\
\hline
 \end{tabular}
 \end{flushleft}
 \end{table*}

 \begin{table*}[htp]
\caption[]{
Total     yields for Z$_{\rm ini}$= 0.020, $\eta_{\rm AGB}$=4, and $m_{\rm HBB}$=0.8  }
 \begin{flushleft}
 \begin{tabular}{lllllllllll}
 \hline
         $m_{\rm ini}$ & H & $^{4}$He & $^{12}$C & $^{13}$C 
       & $^{14}$N & 
  $^{16}$O & Y$_{\rm CNO}$ & Y$_{\rm TOT}$ & 
          $\Delta m_{\rm ej}$ & $m_{\rm end}$ \\ \hline

  .9 & -.992E-02 &  .992E-02 & -.415E-03 &  .309E-04 &  .485E-03 & -.387E-04 &  .616E-04 & -.150E-07 & 
  .33 &   .57 \\ 
 1.0 & -.208E-01 &  .186E-01 & -.123E-03 &  .480E-04 &  .773E-03 &  .981E-03 &  .168E-02 &  .161E-02 & 
  .53 &   .58 \\ 
 1.3 & -.294E-01 &  .257E-01 & -.334E-04 &  .670E-04 &  .122E-02 &  .187E-02 &  .313E-02 &  .303E-02 & 
  .90 &   .59 \\ 
 1.3 & -.157E-01 &  .148E-01 & -.494E-03 &  .491E-04 &  .867E-03 &  .416E-03 &  .839E-03 &  .744E-03 & 
  .72 &   .59 \\ 
 1.5 & -.182E-01 &  .160E-01 &  .804E-03 &  .513E-04 &  .986E-03 &  .371E-03 &  .221E-02 &  .208E-02 & 
  .97 &   .59 \\ 
 1.7 & -.167E-01 &  .143E-01 &  .128E-02 &  .503E-04 &  .102E-02 &  .178E-03 &  .253E-02 &  .237E-02 & 
 1.14 &   .59 \\ 
 2.0 & -.240E-01 &  .200E-01 &  .235E-02 &  .597E-04 &  .127E-02 &  .777E-04 &  .376E-02 &  .394E-02 & 
 1.40 &   .60 \\ 
 2.5 & -.332E-01 &  .276E-01 &  .402E-02 &  .641E-04 &  .154E-02 & -.419E-04 &  .558E-02 &  .560E-02 & 
 1.90 &   .60 \\ 
 3.0 & -.399E-01 &  .329E-01 &  .511E-02 &  .668E-04 &  .173E-02 & -.147E-05 &  .692E-02 &  .692E-02 & 
 2.38 &   .62 \\ 
 4.0 & -.376E-01 &  .313E-01 &  .462E-02 &  .698E-04 &  .182E-02 & -.216E-03 &  .629E-02 &  .624E-02 & 
 3.21 &   .79 \\ 
 5.0 & -.377E-01 &  .314E-01 & -.741E-03 &  .141E-03 &  .729E-02 & -.291E-03 &  .640E-02 &  .628E-02 & 
 4.08 &   .92 \\ 
 7.0 & -.436E-01 &  .358E-01 & -.786E-03 &  .170E-03 &  .976E-02 & -.128E-02 &  .786E-02 &  .774E-02 & 
 5.93 &  1.07 \\ 
 8.0 & -.474E-01 &  .390E-01 & -.923E-03 &  .176E-03 &  .108E-01 & -.160E-02 &  .850E-02 &  .841E-02 & 
 6.85 &  1.15 \\ 
 \hline
 \end{tabular}
 \end{flushleft}
 \end{table*}

 \begin{table*}[htp]
\caption[]{
Pre-AGB   yields for Z$_{\rm ini}$= 0.040, $\eta_{\rm AGB}$=4, and $m_{\rm HBB}$=0.8  }
 \begin{flushleft}
 \begin{tabular}{lllllllllll}
 \hline
         $m_{\rm ini}$ & H & $^{4}$He & $^{12}$C & $^{13}$C 
       & $^{14}$N & 
  $^{16}$O & Y$_{\rm CNO}$ & Y$_{\rm TOT}$ & 
          $\Delta m_{\rm ej}$ & $m_{\rm end}$ \\ \hline

  .8 & -.170E-04 &  .121E-04 &  .119E-05 &  .145E-07 &  .300E-06 &  .257E-05 &  .408E-05 &  .408E-05 & 
  .00 &   .80 \\ 
  .9 & -.151E-04 &  .108E-04 &  .106E-05 &  .129E-07 &  .267E-06 &  .229E-05 &  .363E-05 &  .363E-05 & 
  .00 &   .90 \\ 
 1.0 & -.151E-01 &  .112E-01 &  .790E-03 &  .309E-04 &  .379E-03 &  .201E-02 &  .321E-02 &  .321E-02 & 
  .10 &   .90 \\ 
 1.3 & -.138E-01 &  .101E-01 &  .661E-03 &  .286E-04 &  .432E-03 &  .186E-02 &  .298E-02 &  .298E-02 & 
  .11 &  1.14 \\ 
 1.3 & -.693E-04 &  .506E-04 &  .400E-05 &  .168E-06 &  .170E-05 &  .983E-05 &  .157E-04 &  .157E-04 & 
  .00 &  1.25 \\ 
 1.5 & -.110E-02 &  .799E-03 &  .517E-04 &  .241E-05 &  .389E-04 &  .153E-03 &  .247E-03 &  .247E-03 & 
  .01 &  1.49 \\ 
 1.7 & -.247E-02 &  .179E-02 &  .112E-03 &  .528E-05 &  .918E-04 &  .340E-03 &  .551E-03 &  .551E-03 & 
  .03 &  1.67 \\ 
 2.0 & -.238E-02 &  .175E-02 &  .104E-03 &  .502E-05 &  .924E-04 &  .320E-03 &  .524E-03 &  .524E-03 & 
  .03 &  1.97 \\ 
 2.5 & -.255E-02 &  .189E-02 &  .111E-03 &  .528E-05 &  .102E-03 &  .339E-03 &  .560E-03 &  .560E-03 & 
  .04 &  2.46 \\ 
 3.0 & -.546E-02 &  .406E-02 &  .226E-03 &  .110E-04 &  .223E-03 &  .691E-03 &  .116E-02 &  .116E-02 & 
  .10 &  2.89 \\ 
 4.0 & -.266E-02 &  .200E-02 &  .112E-03 &  .549E-05 &  .110E-03 &  .338E-03 &  .568E-03 &  .568E-03 & 
  .07 &  3.93 \\ 
 5.0 & -.293E-02 &  .219E-02 &  .124E-03 &  .621E-05 &  .121E-03 &  .368E-03 &  .621E-03 &  .621E-03 & 
  .09 &  4.91 \\ 
 7.0 & -.698E-02 &  .517E-02 &  .296E-03 &  .148E-04 &  .290E-03 &  .866E-03 &  .147E-02 &  .147E-02 & 
  .31 &  6.69 \\ 
 \hline
 \end{tabular}
 \end{flushleft}
 \end{table*}

\clearpage

\begin{table*}[thp]
\caption[]{
AGB       yields for Z$_{\rm ini}$= 0.040, $\eta_{\rm AGB}$=4, and $m_{\rm HBB}$=0.8  }
 \begin{flushleft}
 \begin{tabular}{lllllllllll}
 \hline
         $m_{\rm ini}$ & H & $^{4}$He & $^{12}$C & $^{13}$C 
       & $^{14}$N & 
  $^{16}$O & Y$_{\rm CNO}$ & Y$_{\rm TOT}$ & 
          $\Delta m_{\rm ej}$ & $m_{\rm end}$ \\ \hline

  .8 & -.833E-02 &  .833E-02 & -.890E-03 &  .681E-04 &  .104E-02 & -.767E-04 &  .140E-03 &  .484E-07 & 
  .30 &   .53 \\ 
  .9 & -.758E-02 &  .758E-02 & -.863E-03 &  .645E-04 &  .101E-02 & -.733E-04 &  .135E-03 & -.294E-07 & 
  .31 &   .59 \\ 
 1.0 & -.806E-02 &  .806E-02 & -.100E-02 &  .728E-04 &  .117E-02 & -.837E-04 &  .157E-03 & -.761E-07 & 
  .40 &   .60 \\ 
 1.3 & -.763E-02 &  .763E-02 & -.122E-02 &  .818E-04 &  .143E-02 & -.972E-04 &  .189E-03 &  .648E-07 & 
  .58 &   .67 \\ 
 1.5 & -.656E-02 &  .657E-02 & -.142E-02 &  .880E-04 &  .166E-02 & -.108E-03 &  .218E-03 &  .262E-06 & 
  .77 &   .73 \\ 
 1.7 & -.543E-02 &  .543E-02 & -.159E-02 &  .926E-04 &  .186E-02 & -.117E-03 &  .241E-03 &  .983E-07 & 
  .94 &   .76 \\ 
 2.0 & -.906E-02 &  .827E-02 & -.132E-02 &  .116E-03 &  .199E-02 & -.195E-03 &  .595E-03 &  .781E-03 & 
 1.16 &   .80 \\
\hline
 \end{tabular}
 \end{flushleft}
 \end{table*}

\begin{table*}[htp]
\caption[]{
Final AGB yields for Z$_{\rm ini}$= 0.040, $\eta_{\rm AGB}$=4, and $m_{\rm HBB}$=0.8  }
 \begin{flushleft}
 \begin{tabular}{lllllllllll}
 \hline
         $m_{\rm ini}$ & H & $^{4}$He & $^{12}$C & $^{13}$C 
       & $^{14}$N & 
  $^{16}$O & Y$_{\rm CNO}$ & Y$_{\rm TOT}$ & 
          $\Delta m_{\rm ej}$ & $m_{\rm end}$ \\ \hline

  .8 &  .000E+00 &  .000E+00 &  .000E+00 &  .000E+00 &  .000E+00 &  .000E+00 &  .000E+00 &  .000E+00 & 
  .00 &   .53 \\ 
  .9 & -.985E-03 &  .985E-03 & -.112E-03 &  .838E-05 &  .131E-03 & -.952E-05 &  .176E-04 &  .453E-08 & 
  .04 &   .55 \\ 
 1.0 & -.106E-02 &  .106E-02 & -.132E-03 &  .956E-05 &  .154E-03 & -.110E-04 &  .206E-04 & -.396E-08 & 
  .05 &   .55 \\ 
 1.3 & -.151E-02 &  .151E-02 & -.242E-03 &  .162E-04 &  .282E-03 & -.192E-04 &  .373E-04 & -.156E-07 & 
  .11 &   .56 \\ 
 1.5 & -.137E-02 &  .137E-02 & -.297E-03 &  .184E-04 &  .347E-03 & -.226E-04 &  .455E-04 & -.133E-07 & 
  .16 &   .57 \\ 
 1.7 & -.103E-02 &  .103E-02 & -.303E-03 &  .176E-04 &  .353E-03 & -.222E-04 &  .460E-04 &  .274E-09 & 
  .18 &   .58 \\ 
 2.0 & -.273E-02 &  .224E-02 &  .136E-03 &  .207E-04 &  .354E-03 & -.371E-04 &  .474E-03 &  .496E-03 & 
  .21 &   .59 \\
\hline
 \end{tabular}
 \end{flushleft}
 \end{table*}

 \begin{table*}[htp]
\caption[]{
Total     yields for Z$_{\rm ini}$= 0.040, $\eta_{\rm AGB}$=4, and $m_{\rm HBB}$=0.8  }
 \begin{flushleft}
 \begin{tabular}{lllllllllll}
 \hline
         $m_{\rm ini}$ & H & $^{4}$He & $^{12}$C & $^{13}$C 
       & $^{14}$N & 
  $^{16}$O & Y$_{\rm CNO}$ & Y$_{\rm TOT}$ & 
          $\Delta m_{\rm ej}$ & $m_{\rm end}$ \\ \hline

  .9 & -.858E-02 &  .858E-02 & -.974E-03 &  .729E-04 &  .114E-02 & -.805E-04 &  .157E-03 &  .360E-05 & 
  .35 &   .55 \\ 
 1.0 & -.243E-01 &  .204E-01 & -.347E-03 &  .113E-03 &  .171E-02 &  .191E-02 &  .339E-02 &  .321E-02 & 
  .55 &   .55 \\ 
 1.3 & -.229E-01 &  .193E-01 & -.804E-03 &  .127E-03 &  .214E-02 &  .174E-02 &  .321E-02 &  .298E-02 & 
  .80 &   .56 \\ 
 1.3 & -.921E-02 &  .919E-02 & -.146E-02 &  .982E-04 &  .171E-02 & -.107E-03 &  .242E-03 &  .158E-04 & 
  .69 &   .56 \\ 
 1.5 & -.903E-02 &  .874E-02 & -.167E-02 &  .109E-03 &  .205E-02 &  .228E-04 &  .510E-03 &  .247E-03 & 
  .94 &   .57 \\ 
 1.7 & -.893E-02 &  .825E-02 & -.178E-02 &  .115E-03 &  .230E-02 &  .201E-03 &  .838E-03 &  .551E-03 & 
 1.15 &   .58 \\ 
 2.0 & -.142E-01 &  .123E-01 & -.108E-02 &  .142E-03 &  .244E-02 &  .878E-04 &  .159E-02 &  .180E-02 & 
 1.41 &   .59 \\ 
 2.5 & -.223E-01 &  .187E-01 &  .609E-03 &  .152E-03 &  .280E-02 & -.798E-04 &  .348E-02 &  .350E-02 & 
 1.90 &   .60 \\ 
 3.0 & -.318E-01 &  .262E-01 &  .189E-02 &  .158E-03 &  .317E-02 &  .964E-04 &  .532E-02 &  .529E-02 & 
 2.39 &   .61 \\ 
 4.0 & -.388E-01 &  .314E-01 &  .454E-02 &  .161E-03 &  .330E-02 & -.417E-03 &  .758E-02 &  .739E-02 & 
 3.32 &   .68 \\ 
 5.0 & -.452E-01 &  .364E-01 &  .468E-02 &  .168E-03 &  .359E-02 & -.590E-03 &  .785E-02 &  .867E-02 & 
 4.23 &   .77 \\ 
 7.0 & -.460E-01 &  .372E-01 & -.188E-02 &  .276E-03 &  .116E-01 & -.125E-02 &  .873E-02 &  .845E-02 & 
 6.02 &   .98 \\ 
 8.0 & -.513E-01 &  .416E-01 & -.195E-02 &  .277E-03 &  .124E-01 & -.130E-02 &  .947E-02 &  .924E-02 & 
 6.91 &  1.09 \\ 
 \hline
 \end{tabular}
 \end{flushleft}
 \end{table*}

 \begin{table}[thp]
\caption[]{
AGB       yields for Z$_{\rm ini}$= 0.001, $\eta_{\rm AGB}$=4, and $m_{\rm HBB}$=0.9  }
 \begin{flushleft}
 \begin{tabular}{lllll}
 \hline
 M$_{\rm ini}$ & $^{12}$C & $^{13}$C & $^{14}$N & 
      $^{16}$O \\ \hline 

 3.0 &  .379E-02 & -.203E-04 &  .699E-04 &  .335E-03 \\ 
 3.5 &  .420E-02 & -.210E-04 &  .759E-04 &  .365E-03 \\ 
 4.0 &  .474E-02 & -.218E-04 &  .823E-04 &  .408E-03 \\ 
 4.5 &  .197E-02 &  .250E-04 &  .334E-02 &  .426E-03 \\ 
 5.0 &  .472E-03 &  .414E-04 &  .541E-02 &  .338E-03 \\ 
 6.0 &  .571E-03 &  .625E-04 &  .655E-02 &  .355E-04 \\ 
 7.0 &  .613E-03 &  .748E-04 &  .731E-02 &  .465E-04 \\ 
 8.0 &  .603E-03 &  .859E-04 &  .794E-02 &  .960E-04 \\ 
 \hline
 \end{tabular}
 \end{flushleft}
 \end{table}

 \begin{table}[htp]
\caption[]{
Final AGB yields for Z$_{\rm ini}$= 0.001, $\eta_{\rm AGB}$=4, and $m_{\rm HBB}$=0.9  }
 \begin{flushleft}
 \begin{tabular}{lllll}
 \hline
 M$_{\rm ini}$ & $^{12}$C & $^{13}$C & $^{14}$N & 
      $^{16}$O \\ \hline 

 3.0 &  .137E-02 & -.309E-05 &  .984E-05 &  .121E-03 \\ 
 3.5 &  .156E-02 & -.318E-05 &  .105E-04 &  .137E-03 \\ 
 4.0 &  .140E-02 & -.306E-05 &  .108E-04 &  .122E-03 \\ 
 4.5 &  .311E-03 &  .100E-04 &  .899E-03 &  .983E-04 \\ 
 5.0 &  .108E-03 &  .112E-04 &  .116E-02 &  .736E-04 \\ 
 6.0 &  .120E-03 &  .147E-04 &  .130E-02 &  .109E-04 \\ 
 7.0 &  .121E-03 &  .163E-04 &  .137E-02 &  .165E-04 \\ 
 \hline
 \end{tabular}
 \end{flushleft}
 \end{table}

 \begin{table}[htp]
\caption[]{
AGB       yields for Z$_{\rm ini}$= 0.004, $\eta_{\rm AGB}$=4, and $m_{\rm HBB}$=0.9  }
 \begin{flushleft}
 \begin{tabular}{lllll}
 \hline
 M$_{\rm ini}$ & $^{12}$C & $^{13}$C & $^{14}$N & 
      $^{16}$O \\ \hline 

 3.0 &  .583E-02 &  .117E-04 &  .400E-03 &  .380E-03 \\ 
 3.5 &  .393E-02 &  .119E-04 &  .361E-03 &  .243E-03 \\ 
 4.0 &  .402E-02 &  .127E-04 &  .343E-03 &  .264E-03 \\ 
 4.5 &  .159E-02 &  .576E-04 &  .325E-02 &  .274E-03 \\ 
 5.0 &  .215E-03 &  .724E-04 &  .522E-02 &  .194E-03 \\ 
 6.0 &  .284E-03 &  .916E-04 &  .636E-02 & -.166E-03 \\ 
 7.0 &  .297E-03 &  .103E-03 &  .718E-02 & -.242E-03 \\ 
 8.0 &  .248E-03 &  .113E-03 &  .791E-02 & -.316E-03 \\ 
 \hline
 \end{tabular}
 \end{flushleft}
 \end{table}

 \begin{table}[htp]
\caption[]{
Final AGB yields for Z$_{\rm ini}$= 0.004, $\eta_{\rm AGB}$=4, and $m_{\rm HBB}$=0.9  }
 \begin{flushleft}
 \begin{tabular}{lllll}
 \hline
 M$_{\rm ini}$ & $^{12}$C & $^{13}$C & $^{14}$N & 
      $^{16}$O \\ \hline 

 3.0 &  .154E-02 &  .101E-05 &  .358E-04 &  .117E-03 \\ 
 3.5 &  .142E-02 &  .156E-05 &  .483E-04 &  .105E-03 \\ 
 4.0 &  .152E-02 &  .172E-05 &  .473E-04 &  .115E-03 \\ 
 4.5 &  .277E-03 &  .148E-04 &  .907E-03 &  .777E-04 \\ 
 5.0 &  .775E-04 &  .154E-04 &  .118E-02 &  .552E-04 \\ 
 6.0 &  .872E-04 &  .184E-04 &  .132E-02 & -.217E-04 \\ 
 7.0 &  .842E-04 &  .195E-04 &  .140E-02 & -.292E-04 \\ 
 8.0 &  .724E-04 &  .208E-04 &  .147E-02 & -.329E-04 \\ 
 \hline
 \end{tabular}
 \end{flushleft}
 \end{table}

 \begin{table}[htp]
\caption[]{
AGB       yields for Z$_{\rm ini}$= 0.008, $\eta_{\rm AGB}$=4, and $m_{\rm HBB}$=0.9  }
 \begin{flushleft}
 \begin{tabular}{lllll}
 \hline
 M$_{\rm ini}$ & $^{12}$C & $^{13}$C & $^{14}$N & 
      $^{16}$O \\ \hline 

 3.0 &  .537E-02 &  .223E-04 &  .706E-03 &  .185E-03 \\ 
 3.5 &  .351E-02 &  .219E-04 &  .666E-03 &  .628E-04 \\ 
 4.0 &  .343E-02 &  .226E-04 &  .664E-03 &  .568E-04 \\ 
 4.5 &  .179E-02 &  .580E-04 &  .279E-02 &  .732E-04 \\ 
 5.0 & -.521E-04 &  .798E-04 &  .519E-02 &  .211E-05 \\ 
 6.0 & -.781E-05 &  .976E-04 &  .639E-02 & -.428E-03 \\ 
 7.0 & -.173E-04 &  .109E-03 &  .730E-02 & -.599E-03 \\ 
 8.0 & -.101E-03 &  .134E-03 &  .934E-02 & -.926E-03 \\ 
 \hline
 \end{tabular}
 \end{flushleft}
 \end{table}

 \begin{table}[htp]
\caption[]{
Final AGB yields for Z$_{\rm ini}$= 0.008, $\eta_{\rm AGB}$=4, and $m_{\rm HBB}$=0.9  }
 \begin{flushleft}
 \begin{tabular}{lllll}
 \hline
 M$_{\rm ini}$ & $^{12}$C & $^{13}$C & $^{14}$N & 
      $^{16}$O \\ \hline 

 3.0 &  .144E-02 &  .193E-05 &  .626E-04 &  .854E-04 \\ 
 3.5 &  .142E-02 &  .313E-05 &  .966E-04 &  .741E-04 \\ 
 4.0 &  .149E-02 &  .334E-05 &  .993E-04 &  .784E-04 \\ 
 4.5 &  .344E-03 &  .158E-04 &  .877E-03 &  .490E-04 \\ 
 5.0 &  .431E-04 &  .171E-04 &  .123E-02 &  .277E-04 \\ 
 6.0 &  .521E-04 &  .199E-04 &  .140E-02 & -.668E-04 \\ 
 7.0 &  .475E-04 &  .209E-04 &  .148E-02 & -.886E-04 \\ 
 8.0 &  .336E-04 &  .240E-04 &  .172E-02 & -.121E-03 \\ 
 \hline
 \end{tabular}
 \end{flushleft}
 \end{table}

\

\newpage

\

 \begin{table}[htp]
\caption[]{
AGB       yields for Z$_{\rm ini}$= 0.020, $\eta_{\rm AGB}$=4, and $m_{\rm HBB}$=0.9  }
 \begin{flushleft}
 \begin{tabular}{lllll}
 \hline
 M$_{\rm ini}$ & $^{12}$C & $^{13}$C & $^{14}$N & 
      $^{16}$O \\ \hline 

 3.0 &  .391E-02 &  .588E-04 &  .151E-02 & -.235E-03 \\ 
 3.5 &  .351E-02 &  .591E-04 &  .153E-02 & -.283E-03 \\ 
 4.0 &  .318E-02 &  .599E-04 &  .156E-02 & -.329E-03 \\ 
 4.5 &  .309E-02 &  .607E-04 &  .155E-02 & -.334E-03 \\ 
 5.0 &  .302E-02 &  .681E-04 &  .201E-02 & -.337E-03 \\ 
 6.0 & -.744E-03 &  .127E-03 &  .695E-02 & -.944E-03 \\ 
 7.0 & -.805E-03 &  .139E-03 &  .791E-02 & -.132E-02 \\ 
 8.0 & -.957E-03 &  .144E-03 &  .881E-02 & -.171E-02 \\ 
 \hline
 \end{tabular}
 \end{flushleft}
 \end{table}

\vspace{-1.2cm}

 \begin{table}[htp]
\caption[]{
Final AGB yields for Z$_{\rm ini}$= 0.020, $\eta_{\rm AGB}$=4, and $m_{\rm HBB}$=0.9}
 \begin{flushleft}
 \begin{tabular}{lllll}
 \hline
 M$_{\rm ini}$ & $^{12}$C & $^{13}$C & $^{14}$N & 
      $^{16}$O \\ \hline 

 3.0 &  .114E-02 &  .522E-05 &  .136E-03 &  .118E-04 \\ 
 3.5 &  .121E-02 &  .728E-05 &  .190E-03 & -.298E-05 \\ 
 4.0 &  .141E-02 &  .836E-05 &  .219E-03 & -.657E-05 \\ 
 4.5 &  .149E-02 &  .895E-05 &  .231E-03 & -.696E-05 \\ 
 5.0 &  .782E-03 &  .141E-04 &  .633E-03 & -.252E-04 \\ 
 6.0 & -.529E-04 &  .257E-04 &  .164E-02 & -.170E-03 \\ 
 7.0 & -.612E-04 &  .267E-04 &  .175E-02 & -.234E-03 \\ 
 8.0 & -.906E-04 &  .270E-04 &  .188E-02 & -.296E-03 \\ 
 \hline
 \end{tabular}
 \end{flushleft}
 \end{table}

\vspace{-1.2cm}

 \begin{table}[htp]
\caption[]{
AGB       yields for Z$_{\rm ini}$= 0.040, $\eta_{\rm AGB}$=4, and $m_{\rm HBB}$=0.9  }
 \begin{flushleft}
 \begin{tabular}{lllll}
 \hline
 M$_{\rm ini}$ & $^{12}$C & $^{13}$C & $^{14}$N & 
      $^{16}$O \\ \hline 

 3.0 &  .830E-03 &  .130E-03 &  .260E-02 & -.521E-03 \\ 
 3.5 &  .288E-02 &  .134E-03 &  .270E-02 & -.596E-03 \\ 
 4.0 &  .282E-02 &  .136E-03 &  .278E-02 & -.650E-03 \\ 
 4.5 &  .290E-02 &  .140E-03 &  .292E-02 & -.744E-03 \\ 
 5.0 &  .305E-02 &  .142E-03 &  .304E-02 & -.832E-03 \\ 
 6.0 &  .327E-02 &  .147E-03 &  .298E-02 & -.816E-03 \\ 
 7.0 & -.582E-03 &  .204E-03 &  .759E-02 & -.150E-02 \\ 
 8.0 & -.216E-02 &  .212E-03 &  .965E-02 & -.230E-02 \\ 
 \hline
 \end{tabular}
 \end{flushleft}
 \end{table}

\vspace{-1.2cm}

 \begin{table}[htp]
\caption[]{
Final AGB yields for Z$_{\rm ini}$= 0.040, $\eta_{\rm AGB}$=4, and $m_{\rm HBB}$=0.9  }
 \begin{flushleft}
 \begin{tabular}{lllll}
 \hline
 M$_{\rm ini}$ & $^{12}$C & $^{13}$C & $^{14}$N & 
      $^{16}$O \\ \hline 

 3.0 &  .835E-03 &  .171E-04 &  .343E-03 & -.741E-04 \\ 
 3.5 &  .126E-02 &  .171E-04 &  .349E-03 & -.838E-04 \\ 
 4.0 &  .161E-02 &  .199E-04 &  .411E-03 & -.105E-03 \\ 
 4.5 &  .149E-02 &  .194E-04 &  .410E-03 & -.113E-03 \\ 
 5.0 &  .151E-02 &  .198E-04 &  .427E-03 & -.126E-03 \\ 
 6.0 &  .154E-02 &  .195E-04 &  .401E-03 & -.119E-03 \\ 
 7.0 & -.387E-04 &  .369E-04 &  .186E-02 & -.314E-03 \\ 
 8.0 & -.302E-03 &  .394E-04 &  .228E-02 & -.489E-03 \\ 
 \hline
 \end{tabular}
 \end{flushleft}
 \end{table}

 \begin{table*}[thp]
\caption[]{Total AGB yields for H, He for synthetic evolution model}
 \begin{flushleft}
 \begin{tabular}{lllllllllll}
 \hline
  & \multicolumn{2}{c}{Z$_{\rm ini}$=0.001} &
   \multicolumn{2}{c}{Z$_{\rm ini}$=0.004} &
   \multicolumn{2}{c}{Z$_{\rm ini}$=0.008} &
   \multicolumn{2}{c}{Z$_{\rm ini}$=0.020} &
   \multicolumn{2}{c}{Z$_{\rm ini}$=0.040} \\
      $m_{\rm ini}$ & 
      H & $^{4}$He & H & $^{4}$He 
   & H & $^{4}$He & 
  H & $^{4}$He & H & $^{4}$He \\ \hline 

  .8 & -.12E-01 &  .12E-01 & -.11E-01 &  .11E-01 & -.11E-01 &  .11E-01 & -.99E-02 &  .99E-02 & -.83E-02 &  .83E-02 \\ 
  .9 & -.13E-01 &  .13E-01 & -.12E-01 &  .12E-01 & -.12E-01 &  .12E-01 & -.99E-02 &  .99E-02 & -.86E-02 &  .86E-02 \\ 
 1.0 & -.14E-01 &  .14E-01 & -.16E-01 &  .15E-01 & -.13E-01 &  .13E-01 & -.11E-01 &  .11E-01 & -.91E-02 &  .91E-02 \\ 
 1.3 & -.24E-01 &  .21E-01 & -.20E-01 &  .18E-01 & -.18E-01 &  .17E-01 & -.11E-01 &  .11E-01 & -.91E-02 &  .91E-02 \\ 
 1.5 & -.24E-01 &  .21E-01 & -.21E-01 &  .18E-01 & -.22E-01 &  .19E-01 & -.15E-01 &  .13E-01 & -.79E-02 &  .79E-02 \\ 
 1.7 & -.23E-01 &  .20E-01 & -.24E-01 &  .20E-01 & -.22E-01 &  .18E-01 & -.15E-01 &  .13E-01 & -.65E-02 &  .65E-02 \\
\hline
 \end{tabular}
 \end{flushleft}
 \end{table*}

 \begin{table*}[thp]
\caption[]{
AGB       yields for Z$_{\rm ini}$= 0.004, $\eta_{\rm AGB}$=2, and $m_{\rm HBB}$=0.8  }
 \begin{flushleft}
 \begin{tabular}{lllllllllll}
 \hline
         $m_{\rm ini}$ & H & $^{4}$He & $^{12}$C & $^{13}$C 
       & $^{14}$N & 
  $^{16}$O & Y$_{\rm CNO}$ & Y$_{\rm TOT}$ & 
          $\Delta m_{\rm ej}$ & $m_{\rm end}$ \\ \hline

  .8 & -.108E-01 &  .108E-01 & -.753E-04 &  .564E-05 &  .879E-04 & -.649E-05 &  .117E-04 & -.145E-07 & 
  .25 &   .57 \\ 
  .9 & -.113E-01 &  .113E-01 & -.832E-04 &  .609E-05 &  .971E-04 & -.707E-05 &  .129E-04 &  .428E-09 & 
  .30 &   .60 \\ 
 1.0 & -.139E-01 &  .133E-01 &  .415E-03 &  .681E-05 &  .112E-03 &  .338E-04 &  .568E-03 &  .551E-03 & 
  .38 &   .62 \\ 
 1.3 & -.208E-01 &  .182E-01 &  .222E-02 &  .808E-05 &  .145E-03 &  .181E-03 &  .255E-02 &  .253E-02 & 
  .61 &   .64 \\ 
 1.5 & -.273E-01 &  .225E-01 &  .427E-02 &  .902E-05 &  .175E-03 &  .350E-03 &  .480E-02 &  .477E-02 & 
  .87 &   .63 \\ 
 1.7 & -.281E-01 &  .225E-01 &  .502E-02 &  .901E-05 &  .187E-03 &  .411E-03 &  .563E-02 &  .559E-02 & 
 1.02 &   .68 \\ 
 2.0 & -.401E-01 &  .315E-01 &  .757E-02 &  .969E-05 &  .300E-03 &  .571E-03 &  .846E-02 &  .865E-02 & 
 1.28 &   .71 \\
\hline
 \end{tabular}
 \end{flushleft}
 \end{table*}

 \begin{table*}[htp]
\caption[]{
Final AGB yields for Z$_{\rm ini}$= 0.004, $\eta_{\rm AGB}$=2, and $m_{\rm HBB}$=0.8  }
 \begin{flushleft}
 \begin{tabular}{lllllllllll}
 \hline
         $m_{\rm ini}$ & H & $^{4}$He & $^{12}$C & $^{13}$C 
       & $^{14}$N & 
  $^{16}$O & Y$_{\rm CNO}$ & Y$_{\rm TOT}$ & 
          $\Delta m_{\rm ej}$ & $m_{\rm end}$ \\ \hline

  .8 & -.494E-04 &  .494E-04 & -.344E-06 &  .258E-07 &  .402E-06 & -.297E-07 &  .536E-07 & -.967E-10 & 
  .00 &   .57 \\ 
  .9 & -.670E-03 &  .670E-03 & -.493E-05 &  .361E-06 &  .576E-05 & -.419E-06 &  .766E-06 &  .218E-09 & 
  .02 &   .58 \\ 
 1.0 & -.179E-02 &  .152E-02 &  .246E-03 &  .508E-06 &  .839E-05 &  .201E-04 &  .275E-03 &  .273E-03 & 
  .03 &   .59 \\ 
 1.3 & -.265E-02 &  .210E-02 &  .496E-03 &  .564E-06 &  .101E-04 &  .405E-04 &  .547E-03 &  .544E-03 & 
  .05 &   .60 \\ 
 1.5 & -.313E-02 &  .226E-02 &  .806E-03 &  .228E-06 &  .463E-05 &  .659E-04 &  .877E-03 &  .873E-03 & 
  .03 &   .60 \\ 
 1.7 & -.414E-02 &  .302E-02 &  .102E-02 &  .489E-06 &  .104E-04 &  .836E-04 &  .112E-02 &  .111E-02 & 
  .07 &   .61 \\ 
 2.0 & -.552E-02 &  .399E-02 &  .140E-02 &  .449E-06 &  .150E-04 &  .112E-03 &  .152E-02 &  .153E-02 & 
  .08 &   .63 \\
\hline
 \end{tabular}
 \end{flushleft}
 \end{table*}

 \begin{table*}[htp]
\caption[]{
Total     yields for Z$_{\rm ini}$= 0.004, $\eta_{\rm AGB}$=2, and $m_{\rm HBB}$=0.8  }
 \begin{flushleft}
 \begin{tabular}{lllllllllll}
 \hline
         $m_{\rm ini}$ & H & $^{4}$He & $^{12}$C & $^{13}$C 
       & $^{14}$N & 
  $^{16}$O & Y$_{\rm CNO}$ & Y$_{\rm TOT}$ & 
          $\Delta m_{\rm ej}$ & $m_{\rm end}$ \\ \hline

  .9 & -.121E-01 &  .121E-01 & -.853E-04 &  .643E-05 &  .104E-03 & -.652E-06 &  .245E-04 &  .108E-04 & 
  .32 &   .58 \\ 
 1.0 & -.158E-01 &  .150E-01 &  .665E-03 &  .729E-05 &  .122E-03 &  .626E-04 &  .857E-03 &  .839E-03 & 
  .42 &   .59 \\ 
 1.3 & -.235E-01 &  .204E-01 &  .271E-02 &  .863E-05 &  .156E-03 &  .226E-03 &  .311E-02 &  .308E-02 & 
  .66 &   .60 \\ 
 1.3 & -.235E-01 &  .204E-01 &  .271E-02 &  .864E-05 &  .156E-03 &  .226E-03 &  .310E-02 &  .307E-02 & 
  .66 &   .60 \\ 
 1.5 & -.310E-01 &  .253E-01 &  .508E-02 &  .914E-05 &  .190E-03 &  .444E-03 &  .573E-02 &  .569E-02 & 
  .92 &   .60 \\ 
 1.7 & -.327E-01 &  .259E-01 &  .605E-02 &  .940E-05 &  .206E-03 &  .517E-03 &  .678E-02 &  .674E-02 & 
 1.11 &   .61 \\ 
 2.0 & -.459E-01 &  .357E-01 &  .897E-02 &  .101E-04 &  .320E-03 &  .694E-03 &  .100E-01 &  .102E-01 & 
 1.37 &   .63 \\ 
 2.5 & -.539E-01 &  .417E-01 &  .109E-01 &  .114E-04 &  .395E-03 &  .806E-03 &  .121E-01 &  .121E-01 & 
 1.84 &   .66 \\ 
 3.0 & -.575E-01 &  .434E-01 &  .127E-01 &  .116E-04 &  .417E-03 &  .950E-03 &  .141E-01 &  .141E-01 & 
 2.31 &   .69 \\ 
 4.0 & -.405E-01 &  .301E-01 &  .149E-02 &  .396E-03 &  .787E-02 &  .578E-03 &  .103E-01 &  .103E-01 & 
 3.05 &   .95 \\ 
 5.0 & -.492E-01 &  .369E-01 &  .726E-03 &  .145E-03 &  .110E-01 &  .484E-03 &  .123E-01 &  .124E-01 & 
 3.97 &  1.03 \\ 
 7.0 & -.517E-01 &  .382E-01 &  .657E-03 &  .181E-03 &  .128E-01 & -.279E-03 &  .134E-01 &  .135E-01 & 
 5.79 &  1.20 \\ 
 \hline
 \end{tabular}
 \end{flushleft}
 \end{table*}

 \begin{table*}[htp]
\caption[]{
AGB       yields for Z$_{\rm ini}$= 0.001, $\eta_{\rm AGB}$=1, and $m_{\rm HBB}$=0.8  }
 \begin{flushleft}
 \begin{tabular}{lllllllllll}
 \hline
         $m_{\rm ini}$ & H & $^{4}$He & $^{12}$C & $^{13}$C 
       & $^{14}$N & 
  $^{16}$O & Y$_{\rm CNO}$ & Y$_{\rm TOT}$ & 
          $\Delta m_{\rm ej}$ & $m_{\rm end}$ \\ \hline

  .8 & -.118E-01 &  .118E-01 & -.183E-04 & -.104E-04 &  .214E-04 & -.173E-05 & -.911E-05 & -.147E-07 & 
  .27 &   .56 \\ 
  .9 & -.123E-01 &  .123E-01 & -.201E-04 & -.113E-04 &  .234E-04 & -.187E-05 & -.982E-05 & -.153E-07 & 
  .32 &   .58 \\ 
 1.0 & -.168E-01 &  .155E-01 &  .118E-02 & -.126E-04 &  .258E-04 &  .104E-03 &  .129E-02 &  .130E-02 & 
  .39 &   .61 \\ 
 1.3 & -.339E-01 &  .273E-01 &  .603E-02 & -.162E-04 &  .314E-04 &  .534E-03 &  .658E-02 &  .659E-02 & 
  .62 &   .63 \\ 
 1.5 & -.419E-01 &  .325E-01 &  .865E-02 & -.185E-04 &  .362E-04 &  .767E-03 &  .944E-02 &  .945E-02 & 
  .84 &   .66 \\ 
 1.7 & -.426E-01 &  .324E-01 &  .936E-02 & -.193E-04 &  .387E-04 &  .829E-03 &  .102E-01 &  .102E-01 & 
  .99 &   .71 \\ 
 2.0 & -.530E-01 &  .398E-01 &  .120E-01 & -.207E-04 &  .684E-04 &  .105E-02 &  .131E-01 &  .132E-01 & 
 1.23 &   .77 \\
\hline
 \end{tabular}
 \end{flushleft}
 \end{table*}

\begin{table*}[htp]
\caption[]{
Final AGB yields for Z$_{\rm ini}$= 0.001, $\eta_{\rm AGB}$=1, and $m_{\rm HBB}$=0.8  }
 \begin{flushleft}
 \begin{tabular}{lllllllllll}
 \hline
         $m_{\rm ini}$ & H & $^{4}$He & $^{12}$C & $^{13}$C 
       & $^{14}$N & 
  $^{16}$O & Y$_{\rm CNO}$ & Y$_{\rm TOT}$ & 
          $\Delta m_{\rm ej}$ & $m_{\rm end}$ \\ \hline

  .8 & -.108E-03 &  .108E-03 & -.167E-06 & -.952E-07 &  .195E-06 & -.158E-07 & -.831E-07 & -.133E-09 & 
  .00 &   .56 \\ 
  .9 & -.263E-03 &  .263E-03 & -.431E-06 & -.243E-06 &  .503E-06 & -.401E-07 & -.211E-06 & -.322E-09 & 
  .01 &   .58 \\ 
 1.0 & -.235E-02 &  .179E-02 &  .515E-03 & -.657E-06 &  .103E-05 &  .456E-04 &  .560E-03 &  .561E-03 & 
  .02 &   .59 \\ 
 1.3 & -.414E-02 &  .298E-02 &  .106E-02 & -.746E-06 &  .879E-06 &  .941E-04 &  .116E-02 &  .116E-02 & 
  .03 &   .60 \\ 
 1.5 & -.311E-02 &  .220E-02 &  .841E-03 & -.465E-06 &  .447E-06 &  .745E-04 &  .915E-03 &  .915E-03 & 
  .02 &   .64 \\ 
 1.7 & -.519E-02 &  .367E-02 &  .140E-02 & -.904E-06 &  .109E-05 &  .124E-03 &  .152E-02 &  .152E-02 & 
  .04 &   .66 \\ 
 2.0 & -.542E-02 &  .384E-02 &  .144E-02 & -.960E-06 &  .226E-05 &  .128E-03 &  .157E-02 &  .158E-02 & 
  .06 &   .71 \\ 
 2.5 & -.630E-02 &  .439E-02 &  .175E-02 & -.116E-05 &  .289E-05 &  .155E-03 &  .191E-02 &  .191E-02 & 
  .09 &   .79 \\ 
 3.0 & -.405E-02 &  .281E-02 &  .112E-02 &  .482E-06 &  .318E-05 &  .997E-04 &  .123E-02 &  .123E-02 & 
  .11 &   .89 \\ 
 3.5 & -.438E-02 &  .308E-02 &  .804E-03 &  .145E-03 &  .240E-03 &  .105E-03 &  .129E-02 &  .130E-02 & 
  .13 &   .96 \\ 
 4.0 & -.500E-02 &  .357E-02 &  .180E-03 &  .327E-04 &  .111E-02 &  .105E-03 &  .142E-02 &  .143E-02 & 
  .15 &  1.03 \\ 
 4.5 & -.522E-02 &  .373E-02 &  .125E-03 &  .159E-04 &  .124E-02 &  .104E-03 &  .149E-02 &  .149E-02 & 
  .16 &  1.09 \\ 
 5.0 & -.544E-02 &  .389E-02 &  .123E-03 &  .163E-04 &  .134E-02 &  .651E-04 &  .155E-02 &  .156E-02 & 
  .18 &  1.15 \\ 
 \hline
 \end{tabular}
 \end{flushleft}
 \end{table*}

\begin{table*}[htp]
\caption[]{
Total     yields for Z$_{\rm ini}$= 0.001, $\eta_{\rm AGB}$=1, and $m_{\rm HBB}$=0.8  }
 \begin{flushleft}
 \begin{tabular}{lllllllllll}
 \hline
         $m_{\rm ini}$ & H & $^{4}$He & $^{12}$C & $^{13}$C 
       & $^{14}$N & 
  $^{16}$O & Y$_{\rm CNO}$ & Y$_{\rm TOT}$ & 
          $\Delta m_{\rm ej}$ & $m_{\rm end}$ \\ \hline

  .9 & -.132E-01 &  .132E-01 & -.213E-04 & -.128E-04 &  .248E-04 & -.185E-05 & -.111E-04 & -.108E-05 & 
  .36 &   .58 \\ 
 1.0 & -.199E-01 &  .179E-01 &  .169E-02 & -.143E-04 &  .281E-04 &  .150E-03 &  .185E-02 &  .186E-02 & 
  .44 &   .59 \\ 
 1.3 & -.382E-01 &  .305E-01 &  .709E-02 & -.172E-04 &  .328E-04 &  .628E-03 &  .774E-02 &  .775E-02 & 
  .66 &   .60 \\ 
 1.3 & -.385E-01 &  .307E-01 &  .709E-02 & -.176E-04 &  .334E-04 &  .628E-03 &  .774E-02 &  .774E-02 & 
  .67 &   .60 \\ 
 1.5 & -.452E-01 &  .348E-01 &  .949E-02 & -.193E-04 &  .373E-04 &  .841E-03 &  .104E-01 &  .104E-01 & 
  .88 &   .64 \\ 
 1.7 & -.478E-01 &  .361E-01 &  .108E-01 & -.203E-04 &  .400E-04 &  .953E-03 &  .117E-01 &  .117E-01 & 
 1.04 &   .66 \\ 
 2.0 & -.585E-01 &  .437E-01 &  .134E-01 & -.217E-04 &  .710E-04 &  .118E-02 &  .146E-01 &  .148E-01 & 
 1.29 &   .71 \\ 
 2.5 & -.575E-01 &  .413E-01 &  .148E-01 & -.225E-04 &  .769E-04 &  .130E-02 &  .161E-01 &  .163E-01 & 
 1.71 &   .79 \\ 
 3.0 & -.520E-01 &  .365E-01 &  .140E-01 & -.129E-04 &  .697E-04 &  .124E-02 &  .153E-01 &  .155E-01 & 
 2.11 &   .89 \\ 
 4.0 & -.664E-01 &  .479E-01 &  .248E-02 &  .466E-03 &  .141E-01 &  .133E-02 &  .184E-01 &  .185E-01 & 
 2.97 &  1.03 \\ 
 5.0 & -.771E-01 &  .556E-01 &  .169E-02 &  .210E-03 &  .185E-01 &  .958E-03 &  .214E-01 &  .215E-01 & 
 3.85 &  1.15 \\ 
 7.0 & -.148E-01 &  .112E-01 &  .247E-03 &  .364E-04 &  .325E-02 &  .246E-04 &  .356E-02 &  .357E-02 & 
 1.84 &  5.16 \\ 
 \hline
 \end{tabular}
 \end{flushleft}
 \end{table*}

\end{document}